\documentclass[a4paper,UKenglish,cleveref, autoref, thm-restate]{lipics-v2021}

\usepackage[T1]{fontenc}
\usepackage{graphicx}

\usepackage{epsfig}
\usepackage{graphicx}
\usepackage{amssymb}
\usepackage{amsfonts}
\usepackage{amsmath}
\usepackage{verbatim}
\usepackage{url}
\usepackage{stmaryrd}
\usepackage{algorithmic}
\usepackage{algorithm}
\usepackage[all]{xypic}
\usepackage{epic,eepic}
\usepackage{environ}
\usepackage{stmaryrd}
\usepackage{makecell}
\usepackage{tikz}
\usetikzlibrary{arrows,shapes,decorations,automata,backgrounds,fit,matrix}
\usetikzlibrary{positioning}
\usetikzlibrary{snakes}
\usetikzlibrary{petri}
\usepackage{subcaption}
\usepackage{ifpdf}
\ifpdf
\usepackage[framemethod=TikZ]{mdframed}
\usepackage[colorinlistoftodos,prependcaption]{todonotes}

\fi
\nolinenumbers

\newcommand{\nat}{\mathbb{N}}

\newcommand{\tuple}[1]{( #1)}
\newcommand{\set}[1]{\{ #1 \}}

\newcommand{\vect}[1]{\mathbf{#1}}

\newcommand{\Prot}{\mathtt{Pr}}
\newcommand{\RvProt}{\mathtt{RP}}

\newcommand{\trans}{\rightarrow}
    \newcommand{\ltrans}[1]{\xrightarrow{#1}}    
\newcommand{\transP}{\Rightarrow}
\makeatletter
\newcommand{\xRightarrow}[2][]{\ext@arrow 0359\Rightarrowfill@{#1}{#2}}
\makeatother

 \makeatletter
 \newcommand{\problemtitle}[1]{\gdef\@problemtitle{#1}}
\newcommand{\probleminput}[1]{\gdef\@probleminput{#1}}
\newcommand{\problemquestion}[1]{\gdef\@problemquestion{#1}}
\NewEnviron{decproblem}{
  \problemtitle{}\probleminput{}\problemquestion{}
  \BODY
  \par\addvspace{.2\baselineskip}
  \noindent
  \fbox{
    
    \centering
  \begin{tabular}{rl}
    \multicolumn{2}{l}{ \@problemtitle} \\
    \hline
    \rule{0pt}{3ex}
    \normalsize \textbf{Input:} &  \@probleminput \\
    \normalsize \textbf{Question:} &  \@problemquestion
  \end{tabular}
    }
  \par\addvspace{.2\baselineskip}
   
}

\newcommand{\ReachP}{\textsc{Reach}}
\newcommand{\TargetP}{\textsc{Synchro}}
        \newcommand{\FixReachP}[1]{\textsc{FixReach[$#1$]}}    
\newcommand{\FixTargetP}[1]{\textsc{FixSynchro[$#1$]}}

\tikzstyle{pstate}=[rectangle, draw=black, rounded corners]

\tikzstyle{init-left}=[pin={[pin distance=7pt,pin
  edge={<-,black,thick},pin position=left]}]

\tikzstyle{state}=[circle,thick,draw=black,inner sep=0pt,minimum
    size=6mm]
\tikzstyle{finalstate}=[accepting,circle,thick,draw=black,inner sep=0pt,minimum
    size=6mm]

    \tikzstyle{lstate}=[diamond,thick,draw=black,inner sep=0pt,minimum
    size=6mm]
    \tikzstyle{lfinalstate}=[accepting,diamond,thick,draw=black,inner sep=0pt,minimum
    size=6mm]
    
\tikzstyle{statesbis}=[diamond,thick,inner sep=0pt,minimum size=6mm]
\tikzstyle{small-states}=[circle,thick,inner sep=0pt,minimum size=5mm]
\tikzstyle{small-statesbis}=[diamond,thick,inner sep=0pt,minimum size=5mm]
\tikzstyle{init-left}=[pin={[pin distance=7pt,pin edge={<-,black,thick}]left:}]
\tikzstyle{init-right}=[pin={[pin distance=7pt,pin edge={<-,black,thick}]right:}]
\tikzstyle{init-below}=[pin={[pin distance=7pt,pin edge={<-,black,thick}]below:}]
\tikzstyle{vertex}=[double,rounded corners,thick,inner sep=0pt,minimum
size=4mm]
\tikzstyle{vertexbis}=[diamond,double,thick,inner sep=0pt,minimum size=4mm]
\tikzstyle{petri-p}=[circle,thick,inner sep=0pt,minimum size=5mm]
\tikzstyle{petri-t}=[rectangle,thick,inner sep=0pt,minimum width=7mm,minimum height=1mm]
\tikzstyle{petri-t2}=[rectangle,thick,inner sep=0pt,minimum width=1mm,minimum height=7mm]
\tikzstyle{petri-tok}=[circle,inner sep=0pt,minimum size=4pt, color=black,fill=black]
\tikzstyle{gate}=[rectangle,thick,inner sep=0pt,minimum size=6mm,draw=black]

\title{Safety Analysis of Parameterized Networks with Shared Counters}

\author{June Roupin}{IRIF, Universit\'e Paris Cit\'e, France}{}{}{}
\author{Arnaud Sangnier}{DIBRIS, Università di Genova, Italy}{arnaud.sangnier@unige.it}{https://orcid.org/0000-0002-6731-0340}{}

\authorrunning{}
\authorrunning{J. Roupin and A. Sangnier}

\Copyright{}
\Copyright{J. Roupin and A. Sangnier}



\relatedversion{}

\EventEditors{}
\EventNoEds{0}
\EventLongTitle{}
\EventShortTitle{}
\EventAcronym{}
\EventYear{2026}
\EventDate{}
\EventLocation{}
\EventLogo{}
\SeriesVolume{}
\ArticleNo{}

\begin{document}
\maketitle            
 \begin{abstract}
We study networks with an unbounded number of processes sharing
some natural counters that can be decremented, incremented and tested to zero. We look at two safety problems: the first one asks whether there is a network with an execution bringing a process in a given final state and the other one  asks the execution to bring all the processes in the final state. In their full generality, these two problems are undecidable, but we show that decidability can be regained by either using a single counter or by forbidding the use of zero-tests (as it is is the case in Vector Addition Systems with States). For these decidable cases, we obtain low complexity results, the problems being solvable either in non-deterministic logarithmic space or in polynomial time.
\end{abstract}

\section{Introduction}

\textit{Verifying concurrent systems with an unbounded number of participants.} When dealing with the verification of concurrent systems different aspects can be taken into account as the shape of the protocol executed by the different processes, the means of communication and the number of processes. For what concerns this latter aspect, in some cases, the system is designed for an unbounded (but fixed at each peculiar instance) of participants and in order to verify its correctness, one needs to analyze the behavior for all the possible number of participants or to find whether there exists a specific number for which a bug occurs. Such number is hence a parameter of the problem and this is the reason why we refer to this family of problems as \emph{parameterized verification}. One of the difficulty lies in the fact that the number of networks to analyze is infinite. This is the setup that German and Sistla presented in \cite{german-reasoning92}, where they propose algorithms to verify networks of unbounded size where each participant executes the same finite-state protocol and the communication takes place thanks to pairwise rendez-vous, i.e., two processes agree to move synchronously. Interestingly, they show that some non-trivial temporal properties could be checked in polynomial time.


\textit{Varying the means of communication.} In the last two decades, with the interest in developing automatic methods to analyze concurrent systems, many results have been found on the verification of  parameterized  networks considering various means of communication. In \cite{durand-model-fmsd17,esparza-parameterized-acm16}, the authors have studied networks where the entities can write and read values from a finite set into a shared memory. In \cite{aminof-parameterized-vmcai14}, the authors assume that the entities of the network can exchange a token and the  communication primitives are either: give the token or take the token. As explained in \cite{aminof-parameterized-dc18} this latter case is a restriction of the rendez-vous communication. In \cite{esparza-verification-lics99}, the entities communicate by broadcasting messages that can be received by everyone in the network. This setting has then be refined in \cite{delzanno-complexity-concur10,delzanno-complexity-fsttcs12} to consider networks equipped with communication topologies where only the neighbors of a sender can see an emitted message. Closer to concurrent programming language primitives, the authors of \cite{mascle-model-concur23} study the parameterized verification problem for processes sharing locks (with an unbounded number of locks) and in \cite{delzanno-towards-tacas02,guillou-safety-concur23} a different semantics of
rendez-vous has been introduced in order to model the behavior of monitors with their \texttt{wait} and \texttt{signal} primitives. One remark is that in most of these works, except for the one with shared locks where lock creation is allowed, when the number of entities in the system is fixed and each entity runs a finite state protocol, the system to analyze is finite; the difficulty being to deal with an infinite number of instantiations of such systems.

\textit{Manipulating counters.} In this work, we study  parameterized verification problems for networks where each process executes the same finite state program manipulating some shared natural variables (called counters) which are initially set to $0$. Such finite state programs are often referred to as counter machines and they have been intensively studied (in the case where they are executed by a single process). It is well-known that when we allow such programs to manipulate at least two counters and to perform increment, decrement and zero-test (such programs are often called Minsky machines), then even simple verification problems as the reachability of a control state become undecidable \cite{minsky-computation-67}. However decidability can be regained by imposing some restrictions on the counters. As a matter of fact, the reachability of a control state can be solved in \textsc{NL} (non-deterministic logarithmic space) \cite{chistikov-shortest-lmcs19,lafourcade-intruder-rta05} for machines with a single counter (when the updates on the counters are encoded in unary) and other verification problems have been shown to be decidable and solvable with low complexity for this class of systems in \cite{haase-reachability-concur09,goller-model-icalp10}. Forbidding the use of zero-tests leads as well to a class of well studied systems for which many verification problems are decidable, namely the Vector Addition Systems with States (VASS). The control state reachability for VASS is known to be \textsc{EXPspace}-complete \cite{lipton-reach-76,rackoff-covering-78} whereas the reachability of a configuration (where both a control state and a value for the counters is provided) is Ackermann-complete \cite{czerwinski-reachability-focs21,leroux-reachability-focs21,leroux-reachability-lics19}. This latter problem being intractable, some papers have provided  some restrictions to circumvent this issue as for instance \cite{blondin-reachability-acm21,chistikov-tractability-focs24,englert-reachability-lics16}. There is furthermore a strong connection between parameterized verification and VASS: in some cases the behavior of the network can be encoded into a VASS associating to each state of the protocol  a counter whose value represents the number of processes in the associated state. This allows to use the numerous decidable results on VASS to perform parameterized verification see e.g. \cite{german-reasoning92}. However, this translation is not always the best choice in term of complexity.

\textit{Our contributions.} We study two parameterized verification problems for networks of entities with shared counters: the first one, we call the reachability problem (\ReachP), asks whether there is a network in which an execution leads to a configuration where (at least) one process ends in a final state and the second one, called the synchronization problem (\TargetP), can be stated similarly but requires all the processes of the reached configuration to be in the final state. Note that when dealing with a single process, these two problems are equivalent and correspond to control state reachability. We first show that these two problems are undecidable in the general case. Even if this seems straightforward from the fact that the reachability problem is undecidable for two-counter Minsky machines, this first intuition is contradicted by the fact that in our networks it is not easy to extract a single process which can simulate on its own the Minsky machine and we need to rely on a trick. We then show that, as it is the case for classical counter machines, if we restrict our networks to use a single counter or not to use zero-tests, then the two problems become decidable. For the case with a single counter, we show that the reachability problem is \textsc{NL}-complete and that the synchronization problem can be solved in polynomial time (and is \textsc{NL}-complete too when we forbid the use of zero-tests). When we forbid the use of zero-tests, we obtain much better complexity as for VASS, as the two problems can be solved in polynomial time. Note that  our results are not only interesting from a theoretical point of view, as shared counters could be interpreted in the paradigm of concurrent programming languages as semaphores (where a wait decrements the value of the counter and blocks if the value is $0$ and a signal increments it). So our techniques for protocols with no zero-test can be used to verify parameterized systems with semaphores with $0$ as an initial value. Oppositely to the single-process case, the initial value of the counters plays here a crucial role. Indeed, authorizing for instance a counter to have $1$ as an initial value would allow to isolate a process (the first one performing a decrement on this counter) and to simulate afterwards the behavior of a VASS, which would lead to higher complexity bounds for the two considered problems. 

\begin{table}[htbp]
\begin{center}
\begin{tabular}{||c| c| c||} 
 \hline
 & \ReachP & \TargetP \\ [0.5ex] 
 \hline\hline
 No restriction & Undecidable [Thm. \ref{thm:undec}] & Undecidable [Thm. \ref{thm:undec}]\\
 \hline
 One-counter and no zero-test & \textsc{NL}-complete [Thm. \ref{thm:reach-onecounter}] & \textsc{NL}-complete [Thm. \ref{thm:target-one-notest}]  \\
 \hline
 One-counter & \textsc{NL}-complete [Thm. \ref{thm:reach-onecounter}] & in \textsc{Ptime} [Thm. \ref{thm:target-one}]  \\
 \hline
 No zero-test & in \textsc{Ptime} [Thm. \ref{thm:reach-testfree}] & in \textsc{Ptime} [Thm. \ref{thm:target-notest}]  \\ 
 \hline
 \hline
\end{tabular}
\end{center}
\caption{Summary of our results}
\end{table}

\textit{Link with threshold automata.} A model that seems close to the one we consider is the model of threshold automata, see e.g. \cite{kukovec-reachability-concur18,balasubramanian-complexity-atva20}, used to represent the behavior of protocols  designed for  unbounded number of entities communicating via message passing and resilient to failures. Here as well, the protocol is modeled by a finite state automaton where each transition is labeled by some (global) guard and updates of shared variables. The shared variables are counters that  can be only increased, because they are used to simulate the emission of a message whereas the guards can speak about the values of the shared variables but use as well parameters such as the number of failed processes. As a matter of fact, the model is incomparable to the one we study, because the counters behave in a monotonic way, but the guards can express more properties. In \cite{kukovec-reachability-concur18}, the authors have studied an extension where the counters could as well be decremented and they show that reachability is undecidable for this model, but their undecidability proof used the fact that atomically a transition can test for the value of a counter and update it, which is something not allowed in our model.


\section{Networks with shared counters}

\subsection{Preliminaries}

We write $\nat$ to denote the set of natural numbers and $[i,j]$ to
represent the set $\set{k\in \nat \mid i\leq k \mbox{ and } k \leq
  j}$ for $i,j \in \nat$. For a set $E$ and a strictly positive natural $k$, we denote by $E^k$ the set of vectors of size $k$ whose elements are in $E$ and  if $\vect{v} \in E^k$, then  $\vect{v}[i]$ represents the $i^{th}$ element of $\vect{v}$ for all $i \in [1,k]$. In some cases, we might represent a vector $\vect{v} \in E^k$ by the notation $[\vect{v}[1],\vect{v}[2],\ldots,\vect{v}[k]]$. We denote by $\vect{0}$ a vector where all elements are $0$. For a vector $\vect{v} \in E^1$, such that $\vect{v}[1]=e$, we will sometimes use the notation $e$ instead of $[e]$.

\subsection{Counter protocols}

We present now our model of networks. We assume that all
the processes in the network behave
similarly following the same finite-state protocol which manipulates counters taking their values in $\nat$. When following the
protocol, each entity has the ability, in an atomic way, to increase,
decrease or compare the value of a counter with $0$. For a natural $k \in
\nat\setminus\set{0}$, we denote by $ActCount(k)$ the set of actions
over $k$ counters represented by
$\set{\mathtt{inc}(c),\mathtt{dec}(c),\mathtt{zero}(c) \mid 0 < c \leq
  k} \cup \set{\tau}$ where $\mathtt{inc}(c)$ is used to increment the value of the $c^{th}$ counter,
$\mathtt{dec}(c)$ to decrement it and $\mathtt{zero}(c)$ to test if
it is equal to $0$ and $\tau$ corresponds to an internal action having
no effect on the counter values. When we will deal with a single counter, we shall simply write $\mathtt{inc}$, $\mathtt{dec}$  and $\mathtt{zero}$ instead of $\mathtt{inc}(1)$, $\mathtt{dec}(1)$  and $\mathtt{zero}(1)$. 

\begin{definition}[Counter protocol]
A \emph{counter protocol} $\Prot$ is a tuple
$\tuple{Q,k,q_{in},q_f,E}$ where $Q$ is a finite set of
states, $k \in \nat \setminus \set{0}$ is the number of counters, $q_{in} \in Q$ is the initial state, $q_f\in Q$ 
is the final state such that $q_{in} \neq q_f$  and $E
\subseteq Q \times ActCount(k) \times Q$ is the set of edges labeled
by actions over the $k$ counters. 
\end{definition}

A \emph{one-counter protocol} is a counter protocol with a single counter
($k=1$) and a counter protocol is said to be \emph{zero-test
free} if there is no edge labeled with $\mathtt{zero}(c)$ for $0 < c
\leq k$.

Let $\Prot=\tuple{Q,k,q_{in},q_f,E}$ be a counter protocol. A configuration $\gamma$ with $n>0$ processes of  $\Prot$ is a pair
$\gamma=(\vect{s},\vect{v})$ such that  $\vect{s} \in Q^n$ is a vector where $\vect{s}[p]$ represents the state of the $p^{th}$ process for $0 <p \leq n$ and
$\vect{v} \in \nat^k$ where $\vect{v}[c]$ provides the value of
the $c^{th}$ counter for $0 < c \leq k$. We use $|| \gamma||$ to
represent $n$, the number of processes in the configuration $\gamma$,
and denote by $\Gamma_n$ the set of configurations with $n$
processes. The set of configurations is given by $\Gamma=\bigcup_{n \in
  \nat\setminus \set{0}}\Gamma_n$. The set of initial
configurations is equal to
$\Gamma_{in}=\set{\gamma=(\vect{s},\vect{0}) \in \Gamma \mid
  \vect{s}[p]=q_{in} \mbox{ for all } 0 < p \leq  ||\gamma||}$. Hence
in an initial configuration, all processes are in the initial state
$q_{in}$ and all the counter values are set to
$0$. We consider two sets of final configurations: the first one
$\Gamma_{\exists f}=\set{\gamma=(\vect{s},\vect{v}) \in \Gamma \mid
\exists 0<p\leq ||\gamma||.  \vect{s}[p]=q_f }$ corresponds to the
configurations where at least one process is in state $q_f$; in the
second one $\Gamma_{\forall f}=\set{\gamma=(\vect{s},\vect{v}) \in \Gamma \mid
  \vect{s}[p]=q_f \mbox{ for all } 0 < p \leq  ||\gamma||}$ all processes are in state $q_f$. Note that for
the final configurations, the values of the counters are not taken into account.

The network semantics associated to the counter protocol
$\Prot$ is given by the labeled transition
system $(\Gamma,\trans)$ where $\trans \subseteq \Gamma \times \nat
\times E \times \Gamma$ is the transition relation defined as follows. We
have $\gamma \ltrans{p,e} \gamma'$ with $\gamma=(\vect{s},\vect{v})$
and $\gamma'=(\vect{s}',\vect{v}')$ iff $|| \gamma||=||\gamma'||$ and
$0 < p \leq ||\gamma||$ and for all $p' \in [1,||\gamma||]\setminus
\set{p}$, $\vect{s}[p']=\vect{s}'[p']$ and one of the following conditions holds:
\begin{itemize}
\item $e=(\vect{s}[p],\tau,\vect{s}'[p])$ and $\vect{v}=\vect{v}'$;
\item $e=(\vect{s}[p],\mathtt{inc}(c),\vect{s}'[p])$ and
  $\vect{v'}[c]=\vect{v}[c]+1$ and  $\vect{v'}[c']=\vect{v}[c']$ for
  all $c' \in [1,k] \setminus\set{c}$;
\item $e=(\vect{s}[p],\mathtt{dec}(c),\vect{s}'[p])$ and
  $\vect{v'}[c]=\vect{v}[c]-1$ and  $\vect{v'}[c']=\vect{v}[c']$ for
  all $c' \in [1,k] \setminus\set{c}$ (since $\vect{v}' \in \nat^k$
  this implies $\vect{v}[c]>0$);
\item $e=(\vect{s}[p],\mathtt{zero}(c),\vect{s}'[p])$ and
  $\vect{v}[c]=0$ and $\vect{v}=\vect{v}'$.
\end{itemize}
Note that when going from the configuration $\gamma$ to $\gamma'$
the number of processes does not change and at most one process, the one
executing the action given by the edge $e$, changes its state and performs the associated update/test on the counters. Notice that each action is performed atomically as when a process $p$ performs an action, it is not interrupted by
another process. We will write $\gamma \trans \gamma'$ if there exists
$p \in \nat$ and $e \in E$ such that $\gamma \ltrans{p,e} \gamma'$ and
use $\trans^\ast$ to describe the reflexive and transitive closure of
$\trans$. A $n$-process \emph{run} $\rho$ of the counter protocol $\Prot$ is
given by a finite sequence of the form:
$
\rho:=\gamma_0 \ltrans{p_1,e_1} \gamma_1 \ltrans{p_2,e_2}
\gamma_2 \cdots  \ltrans{p_\ell,e_\ell}
\gamma_\ell
$
where $\gamma_0$ is a configuration with $n$ processes. For such a run, we denote by $\mathit{first}(\rho)$ [resp.  $\mathit{last}(\rho)$] its first [resp. last] configuration $\gamma_0$ [resp. $\gamma_\ell$]. We use as well $\mathit{len}(\rho)$ to represent the length $\ell$ of $\rho$ and denote by $\mathit{actions}(\rho)$ the sequence $(p_1,e_1)(p_2,e_2)\ldots(p_\ell,e_\ell)$.  A run $\rho$ is  initial whenever $\mathit{first}(\rho)$ belongs to  $\Gamma_{in}$.
For $i \in [1,\ell]$, we denote by $\rho_{\leq i}$ the prefix $\gamma_0 \ltrans{p_1,e_1} \gamma_1 \ltrans{p_2,e_2}
 \gamma_2 \cdots  \ltrans{p_i,e_i} \gamma_i$  of the run $\rho$.
Given two $n$-process runs $\rho$ and $\rho'$ with $\mathit{last}(\rho)=\mathit{first}(\rho')$, we can build a third run  obtained by executing $\rho'$ after $\rho$ and, which we shall denote as $\rho.\rho'$.

\begin{figure}[htbp]
\centering
\begin{tikzpicture}[node distance=1.7cm]
\tikzstyle{every state}=[inner sep=3pt,minimum size=20pt]
\node(0)[state]{$q_{in}$};

\node(1)[state,above right of=0]{$q_1$}
edge [<-,thick] node[sloped,above] {$\mathtt{inc}(1)$}(0);

\node(2)[state,above of=1]{$q_2$}
edge [<-,thick,bend right] node[left] {$\mathtt{dec}(1)$}(1)
edge [->,thick,bend left] node[right] {$\mathtt{inc}(2)$}(1);ù

\node(3)[state,right of=1]{$q_3$}
edge [<-,thick] node[above] {$\mathtt{dec}(2)$}(1);

\node(4)[state,right of=3]{$q_4$}
edge [<-,thick] node[above] {$\mathtt{dec}(2)$}(3);

\node(5)[state,right of=4]{$q_5$}
edge [<-,thick] node[above] {$\mathtt{zero}(2)$}(4);

\node(6)[state,below right of=5]{$q_f$}
edge [<-,thick] node[above,sloped] {$\mathtt{inc}(3)$}(5);

\node(7)[state,right of=0,xshift=2cm]{$q_6$}
edge [<-,thick] node[below] {$\mathtt{dec}(3)$}(0)
edge [->,thick] node[below] {$\mathtt{zero}(3)$}(6);

\end{tikzpicture}
\caption{Example of a counter protocol}
\label{fig:ex}
\end{figure}
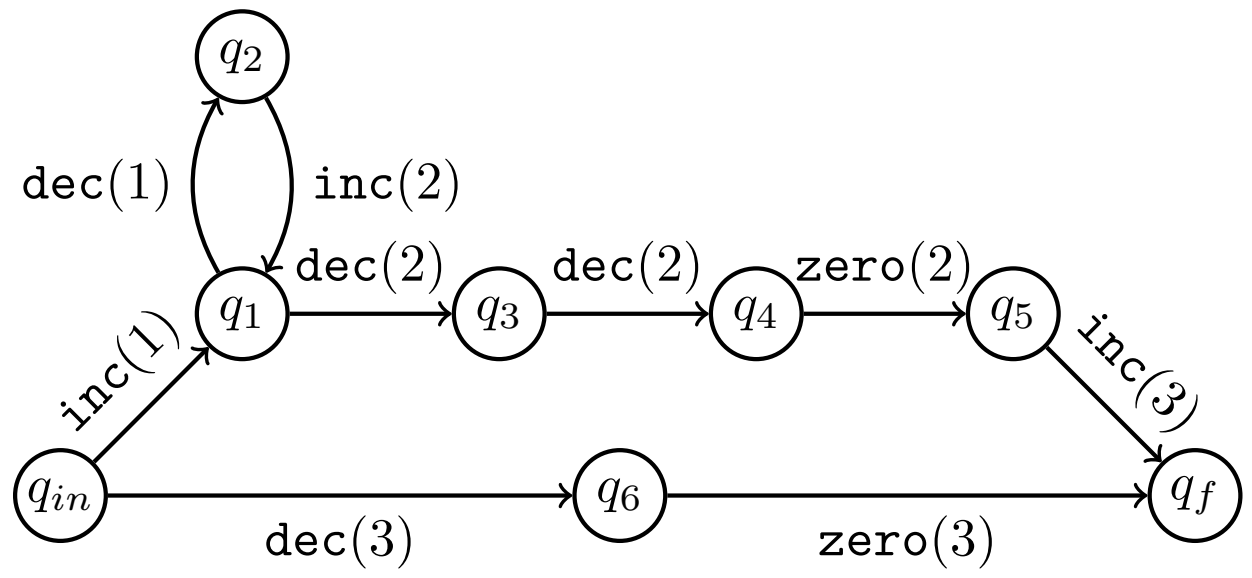

\begin{example}
Figure \ref{fig:ex} presents an example of a counter protocol $\Prot$ with three counters. A possible initial $3$-process run of $\Prot$ is $([q_{in},q_{in},q_{in}],[0,0,0]) \trans ([q_1,q_{in},q_{in}],[1,0,0]) \trans ([q_1,q_1,q_{in}],[2,0,0]) \trans ([q_2,q_1,q_{in}],[1,0,0]) \trans ([q_1,q_1,q_{in}],[1,1,0]) \trans ([q_2,q_1,q_{in}],[0,1,0]) \trans ([q_1,q_1,q_{in}],[0,2,0]) \trans ([q_1,q_3,q_{in}],[0,1,0])  \trans ([q_1,q_4,q_{in}],[0,0,0]) \trans ([q_1,q_5,q_{in}],[0,0,0]) \trans ([q_1,q_f,q_{in}],[0,0,1]) \trans ([q_1,q_f,q_6],[0,0,0]) \trans ([q_1,q_f,q_f],[0,0,0])$. 
\end{example}

\subsection{Verification problems and known results}

In this work, we focus on two parameterized questions which ask whether
there exists a number of processes allowing to reach a configuration
where one process is in the final state or where all processes are in
the final state. The first one called \emph{the reachability problem} is:

\begin{decproblem}
  \problemtitle{\ReachP}
  \probleminput{A counter protocol $\Prot=\tuple{Q,k,q_{in},q_f,E}$;} 
  \problemquestion{Does there exist $\gamma_{in} \in \Gamma_{in}$ and
    $\gamma_f \in \Gamma_{{\exists f}}$ such
    that $\gamma_{in} \trans^\ast \gamma_f$?}
\end{decproblem}

If the answer is positive, we say that $q_f$ is \emph{existentially reachable} by $\Prot$. The second problem is \emph{the synchronization problem}:

\begin{decproblem}
  \problemtitle{\TargetP}
  \probleminput{A counter protocol $\Prot=\tuple{Q,k,q_{in},q_f,E}$;} 
  \problemquestion{Does there exist $\gamma_{in} \in \Gamma_{in}$ and
    $\gamma_f \in \Gamma_{{\forall f}}$ such
    that $\gamma_{in} \trans^\ast \gamma_f$?}
\end{decproblem}

If the answer is positive, we say that $q_f$ is \emph{universally reachable} by $\Prot$. We will as well consider the unparameterized version of these problems
and denote by \FixReachP{n}~[resp. \FixTargetP{n}] (for $n>0$) the variation of \ReachP~[resp. \TargetP] where we impose the
initial configuration to have $n$ processes (i.e., we add the
condition $||\gamma_{in}||=n$) where $n$ is a strictly positive natural encoded in unary. Note that in this
context, \FixReachP{1} and  \FixTargetP{1} are the same  since we have a single process.

The cases where we have a single process are well studied problems and it corresponds to the study of so-called counter machines. When such  machines manipulate at least two counters, then the reachability problem is undecidable \cite{minsky-computation-67}. One way to regain decidability consists in either considering one counter automata or to forbid the use of equality tests with $0$ (and we obtain then VASS). The known results can be reformulated in our context:

\begin{theorem}\label{thm:fixed-result}
  \begin{enumerate}
  \item \FixReachP{1} is undecidable \cite{minsky-computation-67}.
  \item \FixReachP{1} restricted to one-counter protocols is \textsc{NL}-complete \cite{chistikov-shortest-lmcs19,lafourcade-intruder-rta05}.
\item Given two states $q,q'$ of a one-counter protocol, deciding whether $([q],\vect{0}) \trans^\ast ([q]',\vect{0})$ is \textsc{NL}-complete \cite{chistikov-shortest-lmcs19,lafourcade-intruder-rta05}.
  \item \FixReachP{1} restricted to zero-test free protocols is \textsc{EXPspace}-complete \cite{lipton-reach-76,rackoff-covering-78}.
  \end{enumerate}
\end{theorem}

\section{Undecidability in the general case}

We shall see in this section that without any restrictions, \ReachP~and \TargetP~are undecidable. To obtain this result we provide a reduction from \FixReachP{1} with a single process. The main issue to simulate the execution of a single process is that, in the parameterized case, we have to ensure that the executions of different processes 'match' the execution of the single one. This is due to the fact that we cannot isolate a specific process.

\begin{figure}[htbp]
\centering
\begin{subfigure}[b]{\textwidth}
\centering
\begin{tikzpicture}[node distance=1.5cm]
\tikzstyle{every state}=[inner sep=3pt,minimum size=20pt]
\node(0)[state]{$q$};

\node(1)[state,right of=0]{$q'$}
edge [<-,thick] node[above] {$op$}(0);
\end{tikzpicture}
  \caption{An edge $e_i$ from $\Prot$}
\end{subfigure} 
\begin{subfigure}[b]{\textwidth}
\centering
\begin{tikzpicture}[node distance=2.2cm]
\tikzstyle{every state}=[inner sep=3pt,minimum size=20pt]
\node(0)[state]{$q$};

\node(1)[state,right of=0]{$q^i_1$}
edge [<-,thick] node[above] {$\mathtt{dec}(4+i)$}(0);

\node(1bis)[state,right of=1,xshift=-0.6cm]{$q^i_2$}
edge [<-,thick] node[above] {$\mathtt{inc}(4)$}(1);

\node(2)[state,right of=1bis,xshift=0.2cm]{$q^i_3$}
edge [<-,thick] node[above] {$\mathtt{zero}(4+i)$}(1bis);

\node(3)[state,right of=2]{$q^i_4$}
edge [<-,thick] node[above] {$\mathtt{inc}(4+i)$}(2);

\node(3bis)[state,right of=3,xshift=-0.6cm]{$q^i_5$}
edge [<-,thick] node[above] {$\mathtt{dec}(4)$}(3);

\node(3ter)[state,right of=3bis,xshift=-0.4cm]{$q^i_6$}
edge [<-,thick] node[above] {$\mathtt{zero}(4)$}(3bis);

\node(4)[state,right of=3ter,xshift=-1cm]{$q'$}
edge [<-,thick] node[above] {$op$}(3ter);

\end{tikzpicture}
\caption{Its encoding in $\Prot'$}
\end{subfigure}
\caption{Encoding  an edge of $\Prot$}
\label{fig:simul1}
\end{figure}
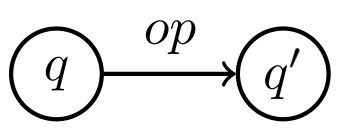
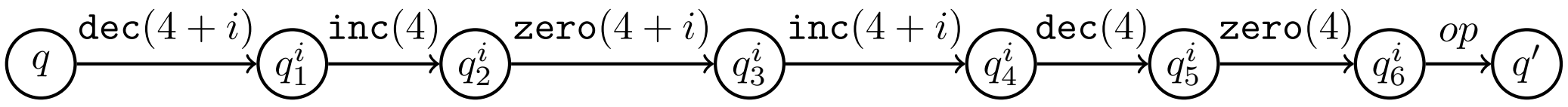

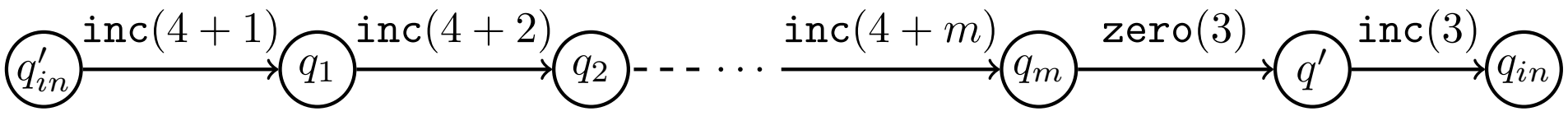
\begin{figure}[htbp]
\centering
\begin{tikzpicture}[node distance=2.2cm]
\tikzstyle{every state}=[inner sep=3pt,minimum size=20pt]
\node(0)[state]{$q'_{in}$};

\node(1)[state,right of=0]{$q_1$}
edge [<-,thick] node[above] {$\mathtt{inc}(4+1)$}(0);

\node(2)[state,right of=1]{$q_2$}
edge [<-,thick] node[above] {$\mathtt{inc}(4+2)$}(1);

\node(3)[right of=2,xshift=-1cm]{$\ldots$}
edge [-,thick,dashed] (2);

\node(4)[state,right of=3,xshift=0.2cm]{$q_m$}
edge [<-,thick] node[above] {$\mathtt{inc}(4+m)$}(3);

\node(5)[state,right of=4]{$q'$}
edge [<-,thick] node[above] {$\mathtt{zero}(3)$}(4);

\node(6)[state,right of=5,xshift=-0.5cm]{$q_{in}$}
edge [<-,thick] node[above] {$\mathtt{inc}(3)$}(5);
\end{tikzpicture}
\caption{Edges of $\Prot'$ for the initialization}
\label{fig:simul2}
\end{figure}

We  present now the key ideas behind our reduction. Let
$\Prot=\tuple{Q,2,q_{in},q_f,E}$ be a counter protocol equipped with two
counters. We assume that $E=\set{e_1,\ldots,e_m}$.  From $\Prot$ we
build another protocol $\Prot'=\tuple{Q',4+m,q'_{in},q_f,E'}$.  It is
equipped with as many counters as edges in
$\Prot$ and with four more counters. The set of states $Q$ of $\Prot$ is included in $Q'$ which contains other auxiliary states depicted on the next figures. The set $E'$ is the smallest set containing the
following edges:
for each transition $e_i \in E$, we have the transitions described by
Figure \ref{fig:simul1} in $E'$ and furthermore we have the edges
provided by Figure \ref{fig:simul2}.

To simulate the execution of $\Prot$ with a single
processes by many processes following $\Prot'$, we have two phases, an
initialization phase and a simulation phase. We recall that in an initial configuration, all counters values are equal to $0$. During the initialization phase, the processes are in the part of
$\Prot'$ described by Figure \ref{fig:simul2} where they consecutively increment for each edge $e_i$  the
$(4+i)^{th}$ counter associated to it. In this initialization phase, the value of
the $(4+i)^{th}$ counter  is equal to the number of
processes that have been through the control state $q_i$.  At the end
of this sequence, the transitions $(q_m,\texttt{zero}(3),q')$
and $(q',\texttt{inc}(3),q_{in})$ are here to ensure some sort of
synchronization. Indeed as soon as a process goes in $q_{in}$ all the
other processes that are not in $q'$ will not be able to go into the
part of the protocol simulating the behavior of $\Prot$.

During the simulation phase, each process  in $q'$ or $q_{in}$ has to simulate the same transitions of $\Prot$ at the same time; this is ensured by
the gadget present at the beginning of the sequences presented in
Figure \ref{fig:simul1}. To take these sequences of
edges, the $(4+i)^{th}$ counter associated to edge $e_i$ should be
decremented to $0$ but since every process involved in the simulation
has incremented it once in the initialization phase, they hence all
have to decrement it, test it with $0$ and reincrement it again (this is ensured by the use of the fourth counter) and finally agree on the same transition (of course some
processes could have incremented these counters and be blocked in the
initialization phase, but in that case, the associated transitions
cannot be simulated). Then the operation of the simulated edge is
performed as many times as there are processes involved in the
simulation. As a matter of fact, if there are $n$ processes involved
in the simulation, if in the actual execution a counter has value
$v$, then in the simulation it has value $n * v$ but this does not
affect the correctness of the simulation as each increment and
decrement will be performed $n$ times. When the value of the
counter is $0$ in the simulated execution, it will be $0$ as well in the simulation. We see then that a single process can reach $q_f$ in $\Prot$ iff there exists an initial configuration $\gamma_{in}$ and a final configuration $\gamma_f \in \Gamma_{\forall f}$ such that $\gamma_{in} \trans^\ast \gamma_f$ in $\Prot'$. Since $\Gamma_{\forall f} \subseteq \Gamma_{\exists f}$ and since if a process can reach $q_f$ in $\Prot'$  then there is an execution where they can all reach $q_f$ this leads to our first result.

\begin{theorem}\label{thm:undec}
  \ReachP~and \TargetP~are undecidable.
\end{theorem}

\begin{remark}
The number of counters used by the protocol we build in our reduction is unbounded as it depends on the number of edges in the original protocol with two counters. We do not know whether the undecidability result still holds for a fixed number of counters. 
\end{remark}

\section{Useful notations}

We introduce here notations useful to analyze the behavior of
counter protocols and their associated network. Let
$\Prot=\tuple{Q,k,q_{in},q_f,E}$ be a counter protocol. A \emph{path}
$P$ in $\Prot$ is either the empty sequence $\varepsilon$ or a finite sequence of edges of the form
$P=(q_0,a_1,q_1)(q_1,a_2,q_2)\linebreak[0]\ldots (q_{m-1},a_m,q_m)$
such that $(q_i,a_i,q_{i+1}) \in E$ for all $i \in [0,m-1]$. For such a non-empty path, we denote by $\mathit{len}(P)$ its length equal to $m$ (the length of an empty path will then be $0$), by
$\mathit{first}(P)$ [resp.$\mathit{last}(P)$]  the first [resp. last] control state $q_0$ [resp. $q_m$] seen on the path and by
$\mathit{lab}(P)$ the finite word of actions $a_1a_2\ldots a_m$ belonging to $ActCount(k)^\ast$. As for runs, we can concatenate two non-empty paths $P$ and $P'$ to obtain a path $P.P'$ whenever $\mathit{last}(P)=\mathit{first}(P')$ (we assume as well $P=\varepsilon.P=P.\varepsilon$).  The value $\mathit{val}_c(P)$ for a counter $c \in [1,k]$ of such a non-empty path $P$  is an integer equal to $|\set{i \in [1,m] \mid a_i=\mathtt{inc}(c)}|-|\set{i \in [1,m] \mid a_i=\mathtt{dec}(c)}|$ (note that this value can be negative). A non-empty path is  initial [resp. final] whenever $\mathit{first}(P)=q_{in}$ [resp. $\mathit{last}(P)=q_{f}$].

Given a $n$-process run 
$\rho:=\gamma_0 \ltrans{p_1,e_1} \gamma_1 \ltrans{p_2,e_2}
\gamma_2 \cdots  \ltrans{p_\ell,e_\ell}
\gamma_\ell$, for all $p \in [1,n]$, we use $\mathit{path}^{(p)}(\rho)$ to denote the path followed by process $p$ in $\rho$. It can be formally defined as follows: it corresponds to the sequence $e_{i_1}\cdots e_{i_k} \in E^\ast$ such that $i_1,\ldots,i_k \in [1,\ell]$ and $p_{i_j}=p$ for all $j \in [1,k]$ and $i_j < i_{j'}$ iff $j < j'$ for all $j,j' \in  [1,\ell]$ and for all $i \in [1,\ell]$, if $p_i=p$ then there exists $j \in [1,k]$ such that $i_j=i$. Note that if $p_i \neq p$ for all $i \in [1,\ell]$ then $\mathit{path}^{(p)}(\rho)=\varepsilon$. A non-empty path $P$ is said to be \emph{feasible with $1$ process} iff there exists a $1$-process run $\rho$ such that $\mathit{first}(\rho)=([\mathit{first}(P)],\vect{0})$  and $\mathit{path}^{(1)}(\rho)=P$.

\section{Solving \ReachP~in some restricted cases}
\label{sec:reach}

In this section, we shall see that when the considered counter protocols are zero-test free, then \ReachP~can be solved in polynomial time and and when these tests are allowed but the protocol manipulates a single counter then \ReachP~is an \textsc{NL}-complete problem.

\subsection{When zero-tests are not allowed}

We begin with the easiest case where we consider a zero-test free counter protocol $\Prot=\tuple{Q,k,q_{in},q_f,E}$. Our method to solve \ReachP~in this case is based on the following observations: first if a counter can reach a strictly positive value at some point then by repeating this run, we can make the value grow as high as we want and second, higher counter values allow to reach more states, since there are no zero-tests.

To formalize these intuitions, for a subset of counters $C \subseteq [1,k]$, we denote by $\Prot_{\setminus C}$ the counter protocol $\tuple{Q,k,q_{in},q_f,E'}$ such that $E'$ is the smallest set verifying:
\begin{itemize}
\item if $(q,a,q') \in E$ and $a \notin \set{\mathtt{inc}(c),\mathtt{dec}(c) \mid c \in C}$ then $(q,a,q') \in E'$, and,
\item if $(q,a,q') \in E$ and $a \in \set{\mathtt{inc}(c),\mathtt{dec}(c) \mid c \in C}$ then $(q,\tau,q') \in E'$.
\end{itemize}
In other words, to obtain $\Prot_{\setminus C}$ from $\Prot$, we replace all updates on counters belonging to $C$ by some silent actions. We have then the following lemma which is at the heart of our decision procedure for \ReachP~in this specific case.

\begin{lemma}\label{lem:reach-zero-test-free}
  $q_f$ is existentially reachable by $\Prot$ if and only if:
  \begin{itemize}
  \item Either there exists a non-empty initial and final path $P$ in $\Prot$ such that $\mathit{lab}(P)$ is in  $\tau^\ast$,
  \item Or, there exists an initial path $P$ in $\Prot$ such that $\mathit{lab}(P)$ belongs to $\tau^\ast\mathtt{inc}(c)$ for some $c \in [1,k]$ and $q_f$ is existentially reachable in $\Prot_{\setminus \{c\}}$.
  \end{itemize}
\end{lemma}

\begin{proof}
  Fist assume $q_f$ is existentially reachable in $\Prot$. We proceed by a case analysis.
  \begin{itemize}
  \item If there does not exist a path $P$ in $\Prot$ such that $\mathit{lab}(P)$ belongs to $\tau^\ast\mathtt{inc}(c)$ for some $c \in [1,k]$, it means that in any run for any counters the first action on counter performed by any process is a decrement, but this is not possible (since in the initial configurations, the initial values of the counters are $0$). So since $q_f$ is existentially reachable in $\Prot$ it means that there is a process following a path  $P$ in $\Prot$ such that $\mathit{lab}(P)$ belongs to $\tau^\ast$ and $\mathit{last}(P)=q_f$.
  \item Assume there exists a path $P$ in $\Prot$ such that $\mathit{lab}(P)$ belongs to $\tau^\ast\mathtt{inc}(c)$ for some $c \in [1,k]$. Then  $q_f$ is existentially reachable in $\Prot_{\setminus \set{c}}$ because all runs that are possible in $\Prot$ are as well feasible in $\Prot_{\setminus \set{c}}$ by ignoring counters $c$ (in this latter protocol we have indeed substituted some actions on counter $c$ by $\tau$).
  \end{itemize}

  Now assume there exists a non-empty path $P$ in $\Prot$ such that $\mathit{lab}(P)$ is in  $\tau^\ast$ and $\mathit{last}(P)=q_f$, then it is clear that $P$ is feasible with $1$ process. Hence $q_f$ is existentially reachable in $\Prot$.

  Finally assume that there exists a path $P$ in $\Prot$ such that $\mathit{lab}(P)$ belongs to $\tau^\ast\mathtt{inc}(c)$ for some $c \in [1,k]$ and $q_f$ is existentially reachable in $\Prot_{\setminus \{c\}}$.  It means that there exists a run in $\Prot_{\setminus \{c\}}$ of the form:
  $
  \rho:=(\vect{s}_0,\vect{v}_0)\ltrans{p_1,e_1} (\vect{s}_1,\vect{v}_1) \ltrans{p_2,e_2}
  (\vect{s}_2,\vect{v}_2) \cdots  \ltrans{p_n,e_n}
  (\vect{s}_n,\vect{v}_n)
  $  where $\vect{v}_0 =\vect{0}$  with $\vect{s}_0[p]=q_{in}$ for all  $0 < p \leq  ||(\vect{s}_0,\vect{v}_0)||$ and such that there exists $0<p\leq ||(\vect{s}_n,\vect{v}_n)||$ verifying $\vect{s}_n[p]=q_f$. When building  $\Prot_{\setminus \{c\}}$ from $\Prot$, each action on the counter $c$ has been replaced by the action $\tau$, we hence deduce that from $\rho$ we can build a run of $\Prot$ which behaves similarly on all the counters except $c$ and where at most $n$ decrements on the counter $c$ has been performed. As a matter of fact, we know that there exists a run in $\Prot $ of the form:
  $
  \rho':=(\vect{s}_0,\vect{v}'_0)\ltrans{p_1,e'_1} (\vect{s}_1,\vect{v}'_1) \ltrans{p_2,e'_2}
  (\vect{s}_2,\vect{v}'_2) \cdots  \ltrans{p_n,e'_n}
  (\vect{s}_n,\vect{v}'_n)
  $
  where $\vect{v}'_0[c]=n$ and for all $i \in [1,n]$ we have :
  \begin{itemize}
  \item $\vect{v}'_i[c']=\vect{v}_i[c']$ for all $c' \in [1,k]\setminus \{c\}$;
  \item $e'_i=e_i$ if $e_i$ is an edge of $\Prot$ and $e'_i \in \set{(q,a,q') \in E \mid a \in \set{\mathtt{inc}(c),\mathtt{dec}(c) \mid c \in C}}$ otherwise.
  \end{itemize}
  Since there exists a path $P$ in $\Prot$ of the form $(q_0,\tau,q_1)(q_1,\tau,q_2)\linebreak[0]\ldots (q_{m-1},\mathtt{inc}(c),q_m)$, we will complete $\rho'$ at the beginning by introducing $n$ new processes all following the path $P$. This introductory step allows us to  reach the configuration $(\vect{s}'_0,\vect{v}'_0)$ such that $||(\vect{s}'_0,\vect{v}'_0)||=N+n$  where $N=||(\vect{s}_0,\vect{v}'_0)||$ and for all $p \in [1,N]$ we have $\vect{s}'_0[p]=q_{in}$ (the $N$ first processes did not move) and for all $p \in [N+1,N+n]$, we have $\vect{s}'[p]=q_m$. From this configuration, following the same actions of $\rho'$ (where the $n$ last processes will not play any role) we can reach the configuration $(\vect{s},\vect{v}'_n)$ where for all $p \in [1,N]$ we have $\vect{s}[p]=\vect{s}_n[p]$  and for all $p \in [N+1,N+n]$, we have $\vect{s}'[p]=q_m$ (the $n$ last processes did not move in these other runs). Since $q_f$ appears in $\vect{s}_n$ we conclude that  $q_f$ is existentially reachable in $\Prot$.
\end{proof}

This last lemma allows us to design a decision procedure to solve \ReachP.  We begin with the zero-test free counter protocol $\Prot=\tuple{Q,k,q_{in},q_f,E}$. If there is a path $P$ in $\Prot$ such that $\mathit{lab}(P)$ is in  $\tau^\ast$ and $\mathit{last}(\rho)=q_f$, we stop and return a positive answer to the problem.  Otherwise we seek for a counter $c \in [1,k]$ for which there exists a path $P$ in $\Prot$ such that $\mathit{lab}(P)$ belongs to $\tau^\ast\mathtt{inc}(c)$, if this is the case, we reinitiate the procedure with the counter-protocol $\Prot_{\setminus \{c\}}$ otherwise we return a negative answer to the problem.

Lemma \ref{lem:reach-zero-test-free} guarantees the correctness of this procedure and since at each step if we do not return a positive answer we delete some actions of a counter $c$, we conclude that we will loop in this procedure at most $k$ times. Hence this algorithm will terminate. Finally, each path search  can be done in non-logarithmic space in the size of the protocol (it corresponds to some reachability query on a graph). Consequently we obtain a polynomial time upper bound for \ReachP~restricted to zero-test free counter protocols.

We furthermore provide a reduction from the Circuit Value Problem, which is \textsc{Ptime}-complete \cite{ladner-circuit-sigact75}, to obtain a matching lower bound. This latter problem consists in determining the value of an output signal in an acyclic boolean circuit when the input values are given. The idea behind the reduction is to use two counters for each value name (or signal) in the circuit, such that if the first [resp. second] counter, called $\top$-counter [resp. $\bot$.counter], is strictly positive, then the signal is evaluated to true [resp. false]. It is then easy to encode the behavior of boolean gates. For instance, to simulate an or signal, if we can decrement one of the two $\top$-counters associated to the input signals, then we increment the $\top$-counter associated to the output and if we can decrement the two $\bot$-counter associated to the inputs, then we increment the $\bot$-counter of the output. Our reduction guarantees that the two counters associated to a signal are never both strictly positive. Note that in this reduction, it does not harm that we have many processes simulating the circuit, as their actions on the different counters will not change the set of counters having a strictly positive value.

As a consequence, we get the following result.

\begin{theorem}\label{thm:reach-testfree}
  \ReachP~restricted to zero-test free counter protocols is \textsc{Ptime}-complete.
\end{theorem}

\begin{proof}
We show that \ReachP~restricted to zero-test free counter protocols is \textsc{Ptime}-hard providing a reduction from \textsc{CVP}, the Circuit Value Problem, which is known to be \textsc{Ptime}-complete \cite{ladner-circuit-sigact75}. This latter problem consists in determining the value of an output signal in an acyclic boolean circuit when the input values are given. An instance of \textsc{CVP} can be defined as a tuple $(\mathit{VN},G)$ where $\mathit{VN}$ is the finite ordered set of value names of form $\set{V_1,V_2,V_3,\ldots,V_n}$ and $G$ is a total function which assigns to each value name a value or a boolean combination of previous values; formally for all $i \in [1,n]$, we have either $G(V_i) \in \set{\top,\bot}$, or $G(V_i)=\neg V_j$ with $j<i$ or $G(V_i)=V_j \vee V_k$ with $j<i$ and $k<i$ or 
$G(V_i)=V_j \wedge V_k$  with $j<i$ and $k<i$. And the \textsc{CVP} asks then to determine whether $V_n$ has a value true ($\top$) following the respective boolean combinations in the circuit.

Let $(\mathit{VN},G)$ be an instance of \textsc{CVP} with $\mathit{VN}=\set{V_1,V_2,V_3,\ldots,V_n}$. From this we build a zero-test free counter protocol $\Prot=\tuple{Q,2n,q_{in},q_f,E}$ with $2n$ counters and where the set $Q$ and $E$ will be described later. Intuitively for each value name $V_i$ the $i^{th}$ [resp. the $(n+i)^{th}$] counter will be assigned a strictly positive value iff $V_i$ will be assigned to true [resp. false]. For each $V_i$, we have the following set of edges in $E$:
\begin{itemize}
\item If $G(V_i)=\top$ then we have the edge $(q_{in},\mathtt{inc}(i),q_i)$ in $E$.
\item If $G(V_i)=\bot$ then we have the edge $(q_{in},\mathtt{inc}(n+i),q_i)$ in $E$.
\item If $G(V_i)=\neg V_j$ then we have the following edges in $E$: $(q_{in},\mathtt{dec}(j),q'_i)$, $(q'_i,\mathtt{inc}(n+i),q_i)$ (these two edges correspond to $V_j$ having true as a value and hence $V_i$ should be set to false), $(q_{in},\mathtt{dec}(n+j),q''_i)$, $(q''_i,\mathtt{inc}(i),q_i)$ (these two edges correspond to $V_j$ having false as a value and hence $V_i$ should be set to true).
\item If $G(V_i)=V_j \wedge V_k$ then we have the following edges in $E$: $(q_{in},\mathtt{dec}(n+j),q'_i)$,$(q_{in},\mathtt{dec}(n+k),q'_i)$ $(q'_i,\mathtt{inc}(n+i),q_i)$ (these last three edge match the case where $V_j$ or $V_k$ are assigned to false, then so  $V_i$ is assigned to false too),  $(q_{in},\mathtt{dec}(j),q''_i)$,$(q''_i,\mathtt{dec}(k),q'''_i)$ $(q'''_i,\mathtt{inc}(i),q_i)$ (these last three edges deal with the case where $V_j$ and $V_k$ are assigned to true).
\item If $G(V_i)=V_j \vee V_k$ then we have the following edges in $E$: $(q_{in},\mathtt{dec}(j),q'_i)$,$(q_{in},\mathtt{dec}(k),q'_i)$ $(q'_i,\mathtt{inc}(i),q_i)$ (these last three edges match the case where $V_j$ or $V_k$ are assigned to true, then $V_i$ is assigned to true too),  $(q_{in},\mathtt{dec}(n+j),q''_i)$,$(q''_i,\mathtt{dec}(n+k),q'''_i)$ $(q'''_i,\mathtt{inc}(n+i),q_i)$ (these last three edges deal with the case where $V_j$ and $V_k$ are assigned to true).
\end{itemize}
And we have as well in $E$ a last edge of the form $(q_{in},\mathtt{dec}(n),q_f)$ which can be taken only when the $n^{th}$ counter has a strictly positive value. The key behind this reduction lies in the fact that since the circuit indeed is acyclic, each time some process takes a sequence of edges to evaluate a value name $V_i$, it will always take the same sequence and finally increase the same associated counter (the $i$-th or the $n+i$-th one). We will finally have that the value name $V_n$ is assigned to true in the circuit iff there exists an initial configuration $\gamma_{in}$ (where all the counters are initially set to $0$) and  a configuration $\gamma_f \in \Gamma_{{\exists f}}$ (where at least of process is in $q_f$ such that $\gamma_{in} \trans^\ast \gamma_f$ (meaning that since the only edge leading to $q_f$ is $(q_{in},\mathtt{dec}(n),q_f)$, then at some point the $n^{th}$ counter has been increased).
\end{proof}

Note that we could have obtained the upper bound thanks to Theorem \ref{thm:target-notest}, which shows that \TargetP~restricted to zero-test free counter protocols is in \textsc{Ptime}. We can indeed reduce, in that case, \ReachP~to \TargetP~(adding a counter that is increased when the final state is reached and transitions from all the other states to the final state decreasing this counter). However, our proof technique provided an \textsc{NL} upper bound when the number of counters is fixed and is a good warm-up for the rest of the paper.

\subsection{Authorizing tests over a single counter}

We consider here a one-counter protocol $\Prot=\tuple{Q,1,q_{in},q_f,E}$. As for the proof of Theorem \ref{thm:reach-testfree}, we provide a necessary and sufficient condition for $q_f$ to be existentially reachable in $\Prot$. We show indeed that if the solution is not trivial, we need an initial and final path  $P_f$ (which will witness the fact that $q_f$ is reachable) and we need two reachable transitions that can increment and decrement the counter on demand (in order to make the path $P_f$ feasible).

\begin{lemma}\label{lem:reach-one}
  $q_f$ is existentially reachable by $\Prot$ if and only if :
  \begin{itemize}
  \item Either the answer to \FixReachP{1}~is true for $\Prot$,
  \item Or, the three following conditions are satisfied:
    \begin{enumerate}
    \item There exists an initial and final path $P_f$ in $\Prot$;
    \item There exists an initial path $P_+$ in $\Prot$ such that $\mathit{lab}(P_+)$ belongs to $(\tau+\mathtt{zero})^\ast\mathtt{inc}$;
    \item There exists a transition $(q_-,\mathtt{dec},q')$ in $E$   and an initial $1$-process run of the form,  $(q_{in},0) \trans^\ast (q_-,0)$.
    \end{enumerate}
  \end{itemize}
\end{lemma}

\begin{proof}
First assume that  $q_f$ is existentially reachable in $\Prot$ and the run to reach a configuration exhibiting $q_f$ needs strictly more than one process (i.e., the answer to \FixReachP{1}~is false for $\Prot$). We have hence a run $\rho$ of the form 
$\gamma_0 \ltrans{p_1,e_1} \gamma_1 \ltrans{p_2,e_2}
\gamma_2 \cdots  \ltrans{p_\ell,e_\ell}
\gamma_\ell$ with $\gamma_i=(\vect{s}_i,v_i)$ for all $i \in [0,\ell]$ where $\gamma_0$ is an initial configuration with $n>1$ processes and $\vect{s}_\ell[1]=q_f$ (we can assume wlog that it is the first process that reaches $q_f$). We prove that the three conditions are satisfied:
\begin{enumerate}
\item By definition of a run and of the transition relation $\trans$, we have that $\mathit{path}^{(1)}(\rho)$ is an initial and final path in $\Prot$.

\item Since, the answer to \FixReachP{1}~is false for $\Prot$, there must be $i \in [1,\ell]$ such that $\mathit{lab}(e_i)=\mathtt{inc}$ (otherwise all for all $i \in [1,\ell]$, we would have $\mathit{lab}(e_i) \in \set{\mathtt{zero},\tau}$ and the answer to \FixReachP{1}~would be true). Let $i_+$ be the first index in $[1,\ell]$ such that $\mathit{lab}(e_{i_+})=\mathtt{inc}$. It means that for all $1 \leq i < i_+$ we have necessarily  $\mathit{lab}(e_i) \in \set{\mathtt{zero},\tau}$ as a decrement of the counter cannot happen if no increment have been seen before. As matter of fact, we deduce that if we take  the path $\mathit{path}^{(p_{i_+})}(\rho_{\leq i_+})$, it corresponds to a path $P_+$ such that $\mathit{lab}(P_+)$ belongs to $(\tau+\mathtt{zero})^\ast\mathtt{inc}$.
\item We let $k$ be the smallest in $[1,\ell]$  index such that $\mathit{path}^{(p_k)}(\rho_{\leq k})$ is not  feasible with $1$ process.  In that case we have necessarily $\mathit{lab}(e_k) \in\set{\mathtt{dec},\mathtt{zero}}$ (the other actions over the counter not blocking a run). We have as well that $\mathit{path}^{(p_k)}(\rho_{\leq (k-1)})$ is feasible with $1$ process and that if $e_k=(q,a,q')$, and if it is not empty we have $\mathit{last}(\mathit{path}^{(p_k)}(\rho_{\leq (k-1)}))=q$ otherwise $q=q_{in}$ (the path followed by process $p_k$ before index $k$ is feasible and it ends in the state $q$, since the next transition taken by $p_k$ is $\rho$ will be $e_k$). We treat the two cases:
  \begin{itemize}
  \item Case $\mathit{lab}(e_k)=\mathtt{dec}$: in that case we have that  $\mathit{val}_1(\mathit{path}^{(p_k)}(\rho_{\leq (k-1)}))=0$. Indeed if $\mathit{val}_1(\mathit{path}^{(p_k)}(\rho_{\leq (k-1)}))>0$ then $\mathit{path}^{(p_k)}(\rho_{\leq k})$ would be feasible (since $k$ is the smallest index for which the path followed by process $p_k$ is not feasible and because of the semantics of the action $\mathtt{dec}$) and if $\mathit{val}_1(\mathit{path}^{(p_k)}(\rho_{\leq (k-1)}))<0$ then $k$ would not be the smallest index for which the path followed by process $p_k$ is not feasible. Hence by putting $e_k=(q_-,\mathtt{dec},q')$ and by taking the run with $1$ process $\rho'$ such that $\mathit{path}^{(p_k)}(\rho_{\leq (k-1)}))=\mathit{path}^{(1)}(\rho')$ and since $\mathit{val}_1(\mathit{path}^{(p_k)}(\rho_{\leq (k-1)}))=0$, we know that the last configuration of $\rho'$ will be of the form  $(q_-,0)$.
  \item Case $\mathit{lab}(e_k)=\mathtt{zero}$: as we have done for the previous case, we first deduce that $\mathit{val}_1(\mathit{path}^{(p_k)}(\rho_{\leq (k-1)}))>0$. Note that by definition of the index $k$ we have as well that $\mathit{val}_1(\mathit{path}^{(p)}(\rho_{\leq (k-1)}))\geq 0$ for all $p \in [1,n]\setminus \set{k}$ (when a path is feasible, the value of the path is the value obtained by the run following the path and is hence a natural). Furthermore by definition of the semantics, we know as well that if $\gamma_{k-1}=(\vect{s}_{k-1},v_{k-1})$ then $v_{k-1}[1]=\Sigma_{p \in [1,n]}\mathit{val}_1(\mathit{path}^{(p)}(\rho_{\leq (k-1)}))$. Consequently, we have $v_{k-1}[1]>0$ which is in contradiction with the fact that $\gamma_{k-1}\ltrans{p_k,e_k} \gamma_{k}$ with $\mathit{lab}(e_k)=\mathtt{zero}$. Hence we cannot have $\mathit{lab}(e_k)=\mathtt{zero}$ and the only valid case is the previous one.
  \end{itemize}
   \end{enumerate} 
  Assume now that the answer to \FixReachP{1}~is true for $\Prot$, then it is clear that $q_f$ is existentially reachable in $\Prot$. So it remains to check that this last fact is satisfied when the three following conditions are satisfied. We first introduce some notations:

  \begin{itemize}
    \item Let $P_f=(q_0,a_1,q_1)(q_1,a_2,q_2)\linebreak[0]\ldots (q_{m-1},a_m,q_m)$
      be an initial and final path in $\Prot$ (hence $q_m=q_f$);
    \item Let $P_+=e_{+,1} \ldots e_{+,m_+}$ be an initial path such $\mathit{lab}(e_{+,i}) \in \set{(\tau,\mathtt{zero}}$ for all $i \in [1,m_+]$ and $\mathit{lab}(e_{+,m_+})=\mathtt{inc}$;
    \item Let $e_{-}=(q_-,\mathtt{dec},q')$ and let  $\rho_{-}$ be the initial $1$ process run of the form 
$\gamma'_0 \ltrans{1,e_{-,1}} \gamma'_1 \ltrans{1,e_{-,2}}
\gamma'_2 \cdots  \ltrans{1,e_{-,m_{-}}}
\gamma_{m_-}$ where $\gamma_{m_-}=(q_-,0)$
  \end{itemize}
  We shall now see how to build $m$ runs $\rho_1,\ldots,\rho_m$ involving $m+1$ processes such that $\rho_i$ will have the form $\gamma_{{i},0} \ltrans{p_{i,1},e_{i,1}} \gamma_{{i},1} \ltrans{p_{i,2},e_{i,2}}
  \gamma_{i,2} \cdots  \ltrans{p_{i,\ell_i},e_{i,\ell_i}}\gamma_{i,\ell_i}$ for all $i \in [1..m]$ and such that $\gamma_{i,\ell_i}=\gamma_{{i+1},0}$ for all $i \in [1..m-1]$ and $\gamma_{{1},0}$ is the initial configuration with $m+1$ processes. In other words, we will be able to 'concatenate' the $m$ runs to obtain an initial run of the form $\gamma_{{1},0} \trans^\ast \gamma_{{1},\ell_1}=\gamma_{{2},0}\trans^\ast \cdots \trans^\ast \gamma_{{m-1},\ell_{m-1}}=\gamma_{{m},0}\trans^\ast\gamma_{{m},\ell_m}$. The shape of each run $\gamma_i$ will be determined by the transition $(q_{i-1},a_i,q_i)$ of $P_f$. Intuitively the fist process will only take this latter transition and the other processes will be here to always set the counter to $0$. We will guarantee in our construction that  if $\gamma_{i,0}=(\vect{s}_{i,0},v_{i,0})$ then $\vect{s}_{i,0}[1]=q_{i-1}$ and $\vect{s}_{i,0}[p]=q_{in}$ for all $p \in [i+1,m+1]$ and $v_{i,0}=0$ and if  $\gamma_{m,\ell_m}=(\vect{s}_{m,\ell_m},v_{m,\ell_m})$ we will have $\vect{s}_{m,\ell_m}[1]=q_m=q_f$. Let $i \in [1,m]$, the run $\rho_i$ has the following form according to $a_i$ the action of transition $e_i=(q_{i-1},a_i,q_i)$:
  \begin{itemize}
  \item \textbf{Case $a_i=\tau$:} we have then $\rho_i:=\gamma_{{i},0} \ltrans{1,e_{i}} \gamma_{{i},1}$ (since our construction guarantees that $\gamma_{i,0}=(\vect{s}_{i,0},0)$ with $\vect{s}_{i,0}[1]=q_{i-1}$);
  \item \textbf{Case $a_i=\mathtt{zero}:$}  we have then $\rho_i:=\gamma_{{i},0} \ltrans{1,e_{i}} \gamma_{{i},1}$ (since our construction guarantees that $\gamma_{i,0}=(\vect{s}_{i,0},0)$ with $\vect{s}_{i,0}[1]=q_{i-1}$, this is possible)
  \item \textbf{Case $a_i=\mathtt{inc}:$} we have then $\rho_i:=\gamma_{{i},0}   \ltrans{1+i,e_{-,1}} \cdots \ltrans{1+i,e_{-,m_{.}}}\ltrans{1,e_{i}}\ltrans{1+i,e_-} \gamma_{{i},m_-+2}$ (the $(i+1)^{th}$ process goes to a point where its effect on the counter is $0$ and ends in the state $q_{-}$, the first process does an increment and the $(i+1)^{th}$ process performs a decrement bringing back the counter to $0$);
\item \textbf{Case $a_i=\mathtt{dec}:$} we have then $\rho_i:=\gamma_{{i},0} \ltrans{1+i,e_{+,1}}\cdots \ltrans{1+i,e_{+,m_{+}}}\gamma_{{i},m_+} \ltrans{1,e_i} \gamma_{{i},m_++1}$ (the $i+1$th process increments the counter and the first process brings it back to $0$).
  \end{itemize}
  One can then show that the concatenation of the runs is possible to obtain finally an initial run ending in the configuration $(\vect{s}_{m,\ell_m},\vect{v}_{m,\ell_m})$ with $\vect{s}_{m,\ell_m}[1]=q_f$. Consequently $q_f$ is existentially reachable in $\Prot$.
\end{proof}

Each of the conditions expressed by the previous lemma can be checked in non-deterministic logarithmic space, either because they correspond to a reachability query on the graph of the protocol or because we can apply the results of Theorem \ref{thm:fixed-result} on  one-counter protocols.

\begin{theorem}\label{thm:reach-onecounter}
  \ReachP~restricted to one-counter protocols is \textsc{NL}-complete.
\end{theorem}

\section{Solving \TargetP~in some restricted cases}
\label{sec:target}

\subsection{No test and one counter}

We will prove here that \TargetP~for zero-test free one-counter protocols can be solved in \textsc{NL}. Again we will rely on some necessary and sufficient conditions on the paths of the considered protocols.
 In the sequel, we fix a zero-test free one-counter protocol $\Prot=\tuple{Q,1,q_{in},q_f,E}$.

\subsubsection{Strictly increasing paths}

We say that a path $P_{inc}$ of the  protocol $\Prot$ is strictly increasing iff it is initial, final and $\mathit{val}_1(P_{inc})>0$. In order to decide in an efficient manner if a protocol has a strictly increasing path, we show that if there exists one, there exists a 'small' one. To obtain this result, we look at the simple cycles in the original strictly increasing path: all the simple cycles with a negative value can be removed and we can keep the first strictly positive cycle to be repeated enough to reach the final state with a strictly positive value.

\begin{lemma}\label{lem:size-strict}
If $\Prot$ has a strictly increasing path, then it has one $P_{inc}$ such that $\mathit{len}(P_{inc}) \leq 2|Q|^2+2|Q|$.
\end{lemma}

\begin{proof}
  Let $P_{inc}$ be a strictly increasing path of $\Prot$. First note, that if $P_{inc}=P.P'.P''$ (with $P$ or $P''$ that could be empty) where $\mathit{last}(P')=\mathit{first}(P')$ and $\mathit{val}_1(P')\leq 0$ then $P.P''$ is a strictly increasing path. We can hence assume that there is no such subpath $P'$ in $P_{inc}$ (they can all be removed). We can now deal with two cases:
  \begin{enumerate}
  \item There is no path $P'$ such that $\mathit{last}(P')=\mathit{first}(P')$ and $P_{inc}=P.P'.P''$. In that case, it means that $P_{inc}$ never goes twice through the same state and as a matter of fact $\mathit{len}(P_{inc}) \leq |Q| \leq 2|Q|^2+2|Q|$.
  \item There is a path $P'$ such that $\mathit{last}(P')=\mathit{first}(P')$ and $P_{inc}=P.P'.P''$. In that case, assume $P_{inc}=(q_0,a_1,q_1)(q_1,a_2,q_2)\linebreak[0]\ldots (q_{m-1},a_m,q_m)$ and let $k$ be the smallest index such there exists $\ell>k$ and $q_k=q_\ell$ and for all $k<i,j\leq \ell$ such that $i \neq j$, we have $q_i \neq q_j$. We denote by $P_{beg}$ the path $(q_0,a_1,q_1)\ldots (q_k,a_k,q_k)$ and by $P_{mid}$ the path (which is a simple cycle) $(q_k,a_{k+1},q_{k+1})(q_{k+1},a_{k+2},q_{k+2})\ldots (q_{\ell-1},a_{\ell},q_{\ell})$. Note that by definition we have $\mathit{len}(P_{beg}) \leq |Q|$ and $\mathit{len}(P_{mid}) \leq |Q|$ and $\mathit{val}_1(P_{mid})> 0$ (since we assumed there is no cycle with a negative value in $P_{inc}$). We consider finally the path $P_{end}=(q_\ell,a_\ell,q_{\ell+1})(q_{\ell+1},a_\ell,q_{\ell+2})\ldots (q_{m-1},a_m,q_m)$. Note that this is a final path. We can modify it such that it does not contain cycle (i.e., subpath $P$ such $\mathit{last}(P)=\mathit{first}(P)$) to obtain another final path $P'_{end}$ such that $ \mathit{first}(P'_{end})=q_\ell$ and  $\mathit{len}(P'_{end}) \leq |Q|$. Note that $P'_{end}$ cannot hence perform more that $|Q|$ decrements and consequently $\mathit{val}_1(P_{end})\geq -|Q|$. We then build the following path $P'_{inc}=P_{beg}(P_{mid})^{2|Q|}P'_{end}$ where $(P_{mid})^{2|Q|}$ means $2|Q|$ repetitions of the simple cycle $P_{mid}$. We have then that $P'_{inc}$ is an initial and final path. We have that $\mathit{len}(P'_{inc}) \leq 2|Q|^2+2|Q|$. And finally, since $\mathit{len}(P_{beg}) \leq |Q|$, we have that  $\mathit{val}_1(P_{beg})\geq -|Q|$ and since $\mathit{val}_1(P'_{inc})=\mathit{val}_1(P_{beg})+2|Q|\mathit{val}_1(P_{mid})+\mathit{val}_1(P'_{end})$ and $\mathit{val}_1(P_{mid})> 0$, we deduce that $\mathit{val}_1(P'_{inc})>0$.
  \end{enumerate}
\end{proof}

Hence to decide if $\Prot$ has a strictly increasing path, we can rely on a \textsc{NL}-algorithm which stores both the length of the path and its value (both can be stored in logarithmic space) and we obtain the matching lower bound from the reachability problem for directed graphs.

\begin{corollary}\label{cor:dec-scrict-path}
  Deciding the existence of a strictly increasing path in a zero-test free one counter protocol is an \textsc{NL}-complete problem.
\end{corollary}

\subsubsection{The protocol has a strictly increasing path}

We begin by providing a condition to solve \TargetP~for protocols with a strictly increasing path. In this case (if there is no trivial initial and final path labeled by $\tau^\ast$) it suffices to have an initial and final path whose first action on the unique counter is an increment. This condition is enough to solve \TargetP, because, we first select  many processes (called the helpers) for taking this first increment to obtain a high enough counter value which enables the strictly increasing path, and then we select other processes to repeatedly take the strictly increasing path to obtain a high enough counter value so that all the helpers can reach the final control state.

\begin{lemma}\label{lem:inc-path-condition}
  Assume $\Prot$ has a strictly increasing path, then $q_f$ is universally reachable by $\Prot$ if and only if:
  \begin{itemize}
  \item Either there exists an initial and final path $P$ in $\Prot$ such that $\mathit{lab}(P)$ is in  $\tau^\ast$ and $\mathit{len}(P)\leq |Q|$,
  \item Or, there exists an initial and final path $P$ in $\Prot$ such that $\mathit{lab}(P)$ belongs to $\tau^\ast\mathtt{inc}\linebreak[0]ActCount(1)^\ast$ and $\mathit{len}(P)\leq 2|Q|$.
  \end{itemize}
\end{lemma}

\begin{proof}
First assume $q_f$ is universally reachable by $\Prot$. It means that there exists  an initial $n$-process run $
\rho:=\gamma_0 \ltrans{p_1,e_1} \gamma_1 \ltrans{p_2,e_2}
\gamma_2 \cdots  \ltrans{p_\ell,e_\ell}
\gamma_\ell
$ such that $\gamma_\ell \in \Gamma_{{\forall f}}$. Note that by definition, we have that $\mathit{path}^{(p)}(\rho)$ is an initial and final path for all $p \in [1,n]$. Assume furthermore there does not exist a non-empty initial and final path $P$ in $\Prot$ such that $\mathit{lab}(P)$ is in  $\tau^\ast$. In that case, the run $\rho$ cannot be composed only of internal actions $\tau$ and since it is a correct run (and the considered protocol is zero-test free), we deduce that there exists $i \in [1,\ell]$ such that $e_i$ is of of the form $(q,\mathtt{inc},q')$ and for all $j \in [1,i-1]$ we have that $e_j$ is of the form $(q'',\tau,q''')$. As a matter of fact $\mathit{lab}(\mathit{path}^{(p_i)}(\rho))$ belongs to $\tau^\ast\mathtt{inc}ActCount(1)^\ast$. Note that furthermore we can reduce the size of this path by deleting the cycles before the action performing the first $\mathtt{inc}(1)$ and as well after. This gives us the bound of $2|Q|$.
Similarly if there exists a non-empty initial and final path $P$ in $\Prot$ such that $\mathit{lab}(P)$ is in  $\tau^\ast$, we can delete all the cycles it takes to obtain a path without cycles of length smaller than $|Q|$.

Assume now that there exists an initial and final path $P$ in $\Prot$ such that $\mathit{lab}(P)$ is in  $\tau^\ast$. Then we can build an initial $1$-process execution $\rho$ such that $\mathit{path}^{(1)}(\rho)=P$ (the single process follows the path $P$ performing only internal actions). In that case we have that $\mathit{last}(\rho)$ belongs to $\Gamma_{{\forall f}}$ (the process ends necessarily in $q_f$ and as a matter of fact $q_f$ is universally reachable by $\Prot$).

Finally, assume there exists an initial and final path $P$ in $\Prot$ such that $\mathit{lab}(P)$ belongs to $\tau^\ast\mathtt{inc}ActCount(1)^\ast$ and $\mathit{len}(P)\leq 2|Q|$. We suppose that $P=P'(q,\mathtt{inc},q_{inc})P''$ where   $\mathit{lab}(P')$ belongs to $\tau^\ast$. Remember that we have assumed that $\Prot$ has a strictly increasing path $P_{inc}$. Let $L_{inc}=\mathit{len}(P_{inc})$. We shall build an initial $L_{inc}+2|Q|L_{inc}$-process run which ends in a configuration where all the processes are in the final state $q_f$. The run begins with a first initial run $\rho_{add}$ where the first $L_{inc}$ processes follow the path $P'$ (which is labeled only by internal actions) and take the transition $(q,\mathtt{inc}(1),q_{inc})$. Formally, we have:
\begin{itemize}
\item $\mathit{first}(\rho_{add}) \in \Gamma_{in}$, and,
\item $\mathit{path}^{(p)}(\rho_{add})=P'(q,\mathtt{inc},q_{inc})$ for all $p \in [1,L_{inc}]$, and,
\item $\mathit{last}(\rho_{add})=(\vect{s}_{add},v_{add})$ where $\vect{s}_{add}[p]=q_{inc}$ for all $p \in [1,L_{inc}]$ and $\vect{s}_{add}[p]=q_{in}$ for all $p \in [L_{inc}+1,L_{inc}+2|Q|L_{inc}]$ and $v_{add}=L_{inc}$ (since the unique operations we have performed on the counter are $L_{inc}$ increments).
\end{itemize}
Now knowing that the path $P_{inc}$ is strictly increasing and that it contains at most $L_{inc}$ decrements, the value of the counter is high enough to make the $2|Q|L_{inc}$ last process follow, one after another the strictly increasing path $P_{inc}$. Indeed each time a process follows this path, the reached counter value always remains strictly bigger than $L_{inc}$, the path being strictly increasing. This gives rise to a new run $\rho_{inc}$ verifying:
\begin{itemize}
\item $\mathit{first}(\rho_{inc})=(\vect{s}_{add},v_{add})$, and,
\item $\mathit{path}^{(p)}(\rho_{inc})=P_{inc}$ for all $p \in [L_{inc}+1,L_{inc}+2|Q|L_{inc}]$, and,
\item $\mathit{last}(\rho_{inc})=(\vect{s}_{inc},v_{inc})$ where $\vect{s}_{inc}[p]=q_{inc}$ for all $p \in [1,L_{inc}]$ (theses processes did not move) and $\vect{s}_{inc}[p]=q_{f}$ for all $p \in [L_{inc}+1,L_{inc}+2|Q|L_{inc}]$ (these processes all followed $P_{inc}$ which is final) and $v_{inc}>L_{inc}+2|Q|L_{inc}$ (the run takes $2|Q|L_{inc}$ times the strictly increasing path and begins with counter value equal to $L_{inc}$).
\end{itemize}
We can now end the path by making the first $L_{inc}$ processes follow the path $P''$ (which ends $P$). This is possible, since this latter path performs at most $2|Q|$ decrements (we recall that $\mathit{len}(P)\leq 2|Q|$). Hence the $L_{inc}$ processes can follow the path without making the counter go below $0$. Formally, we have the run $\rho_{end}$ verifying:
\begin{itemize}
\item $\mathit{first}(\rho_{end})=(\vect{s}_{inc},v_{inc})$, and,
\item $\mathit{path}^{(p)}(\rho_{end})=P''$ for all $p \in [1,L_{inc}]$, and,
\item $\mathit{last}(\rho_{end})=(\vect{s}_{end},v_{end})$ where $\vect{s}_{end}[p]=q_f$ for all $p \in [1,L_{inc}]$ ($P''$ is a final path) and $\vect{s}_{end}[p]=q_{f}$ for all $p \in [L_{inc}+1,L_{inc}+2|Q|L_{inc}]$ (these processes did not move) and $\vect{v}_{end}\geq 0$.
\end{itemize}
By setting $\rho=\rho_{add}.\rho_{inc}.\rho_{end}$, we obtain an initial run where all processes end in state $q_f$.
\end{proof}

\subsubsection{The protocol has no strictly increasing path}

We now give a condition to solve \TargetP~for protocols that do not have a strictly increasing path. In this case (if there is no trivial initial and final path labeled by $\tau^\ast$) it suffices to have two initial and final paths with value $0$, one whose first action on the unique counter is an increment and the other one whose last action on the counter is a decrement.  This condition is enough to solve \TargetP, because, we first select  many processes (the incrementers) for taking this first increment to obtain a high enough counter value which enables the path reaching the decrement. Note that since this latter path has value $0$, then just before reaching the decrement the path has strictly increased the counter value and we can make as many processes as we want (call the decrementers) follow this path in order to let all the incrementers reach the final state. Finally, the path of the incrementers having as well $0$, we show that the decrementers can as well reach the final state. First, we need a technical lemma that states that when there is no strictly increasing path, all the paths followed by processes of an initial run bringing everyone in the final state have value $0$.

\begin{lemma}\label{lem:no-inc-path:zero-val-path}
  Assume $\Prot$ has no strictly increasing path. If $\rho$ is an initial $n$-process run such that $\mathit{last}(\rho) \in \Gamma_{\forall f}$, then $\mathit{val}_1(\mathit{path}^{(p)}(\rho))=0$ for all $p\in [1,n]$ and $\mathit{last}(\rho)=(\vect{s},0)$.
\end{lemma}

\begin{proof}
  Let $\rho$ be an initial $n$-process run and let $\mathit{last}(\rho)=(\vect{s},v)$. Since $\rho$ is a run, we have $v\geq 0$. Note as well that by definition $v=\Sigma_{p \in [1,n]}\mathit{val}_1(\mathit{path}^{(p)}(\rho))$. As $\Prot$ has no strictly increasing path, we know that $\mathit{val}_1(\mathit{path}^{(p)}(\rho)) \leq 0$ for all $p \in [1,n]$. This allows us to deduce that $\Sigma_{p \in [1,n]}\mathit{val}_1(\mathit{path}^{(p)}(\rho))=0$ and that $\mathit{val}_1(\mathit{path}^{(p)}(\rho))=0$ for all $p\in [1,n]$.
\end{proof}

We can now state the necessary and sufficient condition for $q_f$ to be universally reachable in the case $\Prot$ has no increasing path.

\begin{lemma}\label{lem:no-inc-path-condition}
  Assume $\Prot$ has no strictly increasing path, then $q_f$ is universally reachable by $\Prot$ if and only if:
  \begin{itemize}
  \item Either there exists an initial and final path $P$ in $\Prot$ such that $\mathit{lab}(P)$ is in  $\tau^\ast$,
  \item Or, there exists two initial and final paths $P_{inc}$ and $P_{dec}$ in $\Prot$ such that $\mathit{lab}(P_{inc})$ belongs to $\tau^\ast\mathtt{inc}ActCount(1)^\ast$ and $\mathit{lab}(P_{dec})$ belongs to $ActCount(1)^\ast\mathtt{dec}\tau^\ast$ and $\mathit{val}_1(P_{inc})=\mathit{val}_1(P_{dec})=0$ and $\mathit{len}(P_{inc})\leq 2|Q|+1$ and $\mathit{len}(P_{dec})\leq 2|Q|+1$. 
  \end{itemize}
\end{lemma}

\begin{proof}
First assume $q_f$ is universally reachable by $\Prot$. It means that there exists  an initial $n$-process run $
\rho:=\gamma_0 \ltrans{p_1,e_1} \gamma_1 \ltrans{p_2,e_2}
\gamma_2 \cdots  \ltrans{p_\ell,e_\ell} \gamma_\ell$ such that $\gamma_\ell \in \Gamma_{{\forall f}}$. Note that by definition, we have that $\mathit{path}^{(p)}(\rho)$ is an initial and final path for all $p \in [1,n]$. Assume furthermore there does not exist a non-empty initial and final path $P$ in $\Prot$ such that $\mathit{lab}(P)$ is in  $\tau^\ast$. In that case, the run $\rho$ cannot be only composed of internal actions $\tau$ and since it is a correct run (and the considered protocol is zero-test free), we deduce that there exists $i \in [1,\ell]$ such that $e_i$ is of of the form $(q,\mathtt{inc},q')$ and for all $j \in [1,i-1]$ we have that $e_j$ is of the form $(r,\tau,r')$. We denote by $P_{inc}$ the initial and final path $\mathit{path}^{(p_i)}(\rho)$. As a matter of fact $\mathit{lab}(P_{inc})$ belongs to $\tau^\ast\mathtt{inc}ActCount(1)^\ast$. Using Lemma \ref{lem:no-inc-path:zero-val-path}, we furthermore know that $\gamma_\ell=\mathit{last}(\rho)$ is of the form $(\vect{s},0)$, this means that the last action performs on the counter in the run $\rho$ (which is not only composed of internal actions) is a decrement, i.e., there exists $k \in [1,\ell]$ such that $e_k$ is of of the form $(q'',\mathtt{dec},q''')$ and for all $j \in [k+1,\ell]$ we have that $e_j$ is of the form $(r,\tau,r')$. We denote by $P_{dec}$ the initial and final path $\mathit{path}^{(p_k)}(\rho)$ which verifies $\mathit{lab}(P_{dec})$ belongs to $ActCount(1)^\ast\mathtt{dec}\tau^\ast$. Thanks to Lemma \ref{lem:no-inc-path:zero-val-path}, we have as well $\mathit{val}_1(P_{inc})=\mathit{val}_1(P_{dec})=0$.

We now prove that we can reduce the lengths of $P_{inc}$ and $P_{dec}$ to the desired one. Assume $P_{inc}=P_1(q,\mathtt{inc},q')P_2$ with $\mathit{lab}(P_1)$ belonging to $\tau^\ast$. First note that if the path $P_1$ goes twice through the same state ,we can remove the cycle (labeled by internal actions) without changing the value of $P_{inc}$. Hence there exists a path $P'_{inc}=P'_1(q,\mathtt{inc},q')P_2$ with $\mathit{len}(P'_1)<|Q|$ and $\mathit{lab}(P'_1)$ belongs to $\tau^\ast$. Assume now that $P_2=P'_2.P'_3.P'_4$ with $\mathit{first}(P'_3)=\mathit{last}(P'_3)$ (i.e., $P'_3$ is a cycle). In that case, we cannot have $\mathit{val}_1(P'_3)>0$, otherwise by iterating it a big enough number of times, we could obtain from $P'_{inc}$ a strictly increasing path. Note that it is as well not possible that $\mathit{val}_1(P'_3)<0$, because otherwise by removing this cycle from $P'_{inc}$ (which as value $0$) we would get a strictly increasing path. Consequently, $\mathit{val}_1(P'_3)=0$ and we can remove this cycle from $P'_{inc}$ without changing its value. We can repeat this operation until there is no more repeated states in $P_2$, this leads to a path $P''_{inc}=P'_1(q,\mathtt{inc},q')P''_2$ where $\mathit{len}(P''_2)< |Q|$. Consequently $P''_{inc}$ is an initial and final path such that $\mathit{lab}(P''_{inc})$ belongs to $\tau^\ast\mathtt{inc}ActCount(1)^\ast$ and $\mathit{val}_1(P''_{inc})=0$ and $\mathit{len}(P''_{inc})\leq 2|Q|$. Similarly, from $P_{dec}$, we can build an initial and final path $P''_{dec}$ such that $\mathit{lab}(P''_{dec})$ belongs to $ActCount(1)^\ast\mathtt{dec}\tau^\ast$ and $\mathit{val}_1(P''_{dec})=0$ and $\mathit{len}(P''_{dec})\leq 2|Q|$.

As for the case with a strictly increasing path, if there exists a non-empty initial and final path $P$ in $\Prot$ such that $\mathit{lab}(P)$ is in  $\tau^\ast$, then we can build an initial $1$ process execution $\rho$ such that $\mathit{path}^{(1)}(\rho)=P$ and $\mathit{last}(\rho)$ belongs to $\Gamma_{{\forall f}}$.

Finally, assume there exist two initial and final paths $P_{inc}$ and $P_{dec}$ in $\Prot$ such that $\mathit{lab}(P_{inc})$ belongs to $\tau^\ast\mathtt{inc}ActCount(1)^\ast$ and $\mathit{lab}(P_{dec})$ belongs to $ActCount(1)^\ast\mathtt{dec}\tau^\ast$ and $\mathit{val}_1(P_{inc})=\mathit{val}_1(P_{dec})=0$. We suppose that $P_{inc}=P_1(q,\mathtt{inc},q_{inc})P_3$ and that $P_{dec}=P_2(q_{dec},\mathtt{dec},q')P_4$ where $\mathit{lab}(P_1)$ and $\mathit{lab}(P_4)$ belongs to $\tau^\ast$. Note that since $\mathit{val}_1(P_{inc})=0$ then $\mathit{val}_1(P_3)=-1$ and since $\mathit{val}_1(P_{dec})=0$, then $\mathit{val}_1(P_2)=1$.  Let $L_{inc}=\mathit{len}(P_{inc})$ and $L_{dec}=\mathit{len}(P_{dec})$. We show how to build an initial $L_{dec}+L_{inc}$-process run which ends in a configuration in $\Gamma_{\forall f}$. The run begins with a first initial run $\rho_{inc}$ where the first $L_{dec}$ follows the path $P_1(q,\mathtt{inc},q_{inc})$ (remember that $P_1$ is labeled only with internal actions). Formally, we have:
\begin{itemize}
\item $\mathit{first}(\rho_{inc}) \in \Gamma_{in}$, and,
\item $\mathit{path}^{(p)}(\rho_{inc})=P_1(q,\mathtt{inc},q_{inc})$ for all $p \in [1,L_{dec}]$, and,
\item $\mathit{last}(\rho_{inc})=(\vect{s}_{inc},v_{inc})$ where $\vect{s}_{inc}[p]=q_{inc}$ for all $p \in [1,L_{dec}]$ and $\vect{s}_{inc}[p]=q_{in}$ for all $p \in [L_{dec}+1,L_{dec}+L_{inc}]$ and $v_{inc}=L_{dec}$ (since the unique operations we have performed on the counter are $L_{dec}$ increments).
\end{itemize}
Now knowing that $P_2$ is an initial path whose increase strictly the counter value and that performs at most $L_{dec}$ decrements, the value of the counter is high enough to make the $L_{inc}$ last processes follow, one after each other the  path $P_{2}$ (which ends in state $q_{dec}$). Indeed each time a process follows this path, the reached counter value always remains strictly bigger than $L_{dec}$. This gives rise to a new run $\rho_{dec}$ verifying:
\begin{itemize}
\item $\mathit{first}(\rho_{dec})=(\vect{s}_{inc},v_{inc})$, and,
\item $\mathit{path}^{(p)}(\rho_{dec})=P_{2}$ for all $p \in [L_{dec}+1,L_{dec}+L_{inc}]$, and,
\item $\mathit{last}(\rho_{dec})=(\vect{s}_{dec},v_{dec})$ where $\vect{s}_{dec}[p]=q_{inc}$ for all $p \in [1,L_{dec}]$ (theses processes did not move) and $\vect{s}_{dec}[p]=q_{dec}$ for all $p \in [L_{dec}+1,L_{dec}+L_{inc}]$ (these processes all followed $P_{2}$) and $v_{dec}=L_{dec}+L_{inc}$ (the run takes $L_{inc}$ times the path $P_2$ and begins with counter value equal to $L_{dec}$).
\end{itemize}
The next step consists in bringing the first $L_{dec}$ processes in the final state following $P_3$. We know that this path performs at most $L_{inc}$ decrements and at the end it has decreased the counter value of $1$. As before, we can build a third run $\rho'_{inc}$ verifying:
\begin{itemize}
\item $\mathit{first}(\rho'_{inc})=(\vect{s}_{dec},v_{dec})$, and,
\item $\mathit{path}^{(p)}(\rho'_{inc})=P_{3}$ for all $p \in [1,L_{dec}]$, and,
\item $\mathit{last}(\rho'_{inc})=(\vect{s}'_{inc},v'_{inc})$ where $\vect{s}'_{inc}[p]=q_f$ for all $p \in [1,L_{dec}]$ (theses processes follow the final path $P_3$) and $\vect{s}'_{inc}[p]=q_{dec}$ for all $p \in [L_{dec}+1,L_{dec}+L_{inc}]$ (these processes did not move) and $v'_{inc}=L_{inc}$ (the run takes $L_{dec}$ times the path $P_3$ and begins with counter value equal to $L_{dec}+L_{inc}$).
\end{itemize}
The last step consists in bringing the last $L_{inc}$ which are in states $q_{dec}$ in the final state following the path $(q_{dec},\mathtt{dec},q')P_4$, since $P_4$ consists only of internal actions, it means we need to be able to perform $L_{inc}$ decrements, which is feasible as the value of the counter in  $(\vect{s}'_{inc},v'_{inc})$ is $L_{inc}$. We hence can build a fourth run $\rho'_{dec}$ verifying:
\begin{itemize}
\item $\mathit{first}(\rho'_{dec})=(\vect{s}'_{inc},v'_{inc})$, and,
\item $\mathit{path}^{(p)}(\rho'_{dec})=(q_{dec},\mathtt{dec},q')P_4$ for all $p \in [L_{dec}+1,L_{dec}+L_{inc}]$, and,
\item $\mathit{last}(\rho'_{dec})=(\vect{s}'_{dec},v'_{dec})$ where $\vect{s}'_{dec}[p]=q_f$ for all $p \in [1,L_{dec}]$ (theses processes do not move) and $\vect{s}'_{dec}[p]=q_f$ for all $p \in [L_{dec}+1,L_{dec}+L_{inc}]$ (these processes follow the final path $(q_{dec},\mathtt{dec},q')P_4$) and $v'_{dec}=0$ (the run takes $L_{inc}$ times the path $(q_{dec},\mathtt{dec},q')P_4$ and begins with counter value equal to $L_{inc}$).
\end{itemize}

By setting $\rho=\rho_{inc}.\rho_{dec}.\rho'_{inc}.\rho'_{dec}$, we obtain an initial run where all processes end in state $q_f$.
\end{proof}

\subsubsection{Result}

Thanks to Corollary \ref{cor:dec-scrict-path}, we have an \textsc{NL} algorithm to know whether we are in the case where a zero-test free one counter protocol has a strictly increasing path. In the positive case, using Lemma \ref{lem:inc-path-condition}, we can design an \textsc{NL} algorithm for the synchronization problem. Indeed, we can check the sufficient and necessary condition given by the lemma,  looking for paths of length at most twice the number of states and whose counter value is hence bounded as well by this value which can be stored in logarithmic space. The same reasoning holds for the negative case relying on Lemma \ref{lem:no-inc-path-condition}. Here again we obtain the matching lower bound from the reachability problem in directed graph.

\begin{theorem}\label{thm:target-one-notest}
\TargetP~restricted to  zero-test free one-counter protocols is \textsc{NL}-complete.
\end{theorem}

\subsection{A detour to rendez-vous networks}

A well-studied model of parameterized networks is the one where the different processes communicate thanks to pairwise rendez-vous, a process sends a message $m$ that has to be received simultaneously by another process (otherwise the sender cannot move). This model has been introduced in \cite{german-reasoning92} and studied for instance in \cite{aminof-parameterized-dc18,balasubramanian-finding-fossacs21,horn-deciding-concur20}. We shall see that we can use results obtained for this model to obtain complexity results for counter protocols.

For a finite set of messages $\Sigma$, we denote by $ActRdv(\Sigma)$ the set of rendez-vous actions
over $\Sigma$ represented by $\set{!m,?m} \cup \set{\tau}$ where $!m$ [resp. $?m$] represents the request [resp. response] of a rendez-vous over the message $m$ and $\tau$ corresponds to an internal action.

\begin{definition}[Rendez-vous protocol]
A \emph{rendez-vous protocol} $\RvProt$ is a tuple
$\tuple{Q,\Sigma,q_{in},q_f,E}$ where $Q$ is a finite set of
states, $\Sigma$ is a finite alphabet of messages, $q_{in} \in Q$ is the initial state, $q_f\in Q$ 
is the final state such that $q_{in} \neq q_f$  and $E
\subseteq Q \times ActRdv(\Sigma) \times Q$ is the set of edges labeled
by rendez-vous actions over $\Sigma$. 
\end{definition}

A configuration $\xi$ with $n>0$ processes of  $\RvProt$ is a vector $Q^n$ whose size $n$ is denoted by $||\xi||$. The set of configurations is given by $\Xi=\bigcup_{n \in \nat\setminus \set{0}}Q^n$ . The set of initial
configurations is equal to
$\Xi_{in}=\set{\xi \in Q^n \mid
  \vect{\xi}[p]=q_{in} \mbox{ for all } 0 < p \leq  ||\xi||}$. We denote by $\Xi_{\forall f}$ the set $\set{\xi \in Q^n \mid
  \vect{\xi}[p]=q_f \mbox{ for all } 0 < p \leq  ||\xi||}$ where  all processes are in state $q_f$. 

The network semantics associated to a rendez-vous protocol
$\RvProt=\tuple{Q,\Sigma,q_{in},q_f,E}$ is given by the transition
system $(\Xi,\transP)$ where $\transP \subseteq \Xi \times \Xi$ is the transition relation defined as follows. We
have $\xi \transP \xi'$ iff $||\xi||=||\xi'||$ and one of the following conditions holds:
\begin{itemize}
\item there exists $p \in [1,||\xi||]$ such that $(\xi[p],\tau,\xi'[p]) \in E$ and $\vect{\xi}[p']=\vect{\xi}'[p']$ for all $p' \in [1,||\xi||]\setminus\set{p}$ (one process performs an internal action).
\item there exists $p,p' \in [1,||\xi||]$ and $m \in \Sigma$ such that $(\vect{\xi}[p],!m,\vect{\xi}'[p]),(\vect{\xi}[p'],?m,\vect{\xi}'[p']) \in E$ and $\vect{\xi}[p'']=\vect{\xi}'[p'']$ for all $p'' \in [1,||\xi||]\setminus\set{p,p'}$ (two processes synchronize with a rendez-vous over message $m$). 
\end{itemize}
We denote then by $\transP^\ast$ the reflexive and transitive closure of $\transP$. As for counter protocols, we say that the final state is universally reachable by the rendez-vous protocol $\RvProt$ iff there exists an initially configuration $\xi_{in} \in \Xi_{in}$ and a configuration $\xi_f \in \Xi_{\forall f}$ such that $\xi_{in} \transP^\ast \xi_{f}$.

In \cite{balasubramanian-finding-fossacs21}, the authors study the cut-off problem for rendez-vous protocols, which consists in deciding whether there exists a bound $B \in \nat$ such that for all $m \geq B$ and for all initial configurations $\xi_{in} \in Q^m$, there exists a configuration $\xi_f \in \Xi_{\forall f}$ verifying $\xi_{in} \transP^\ast \xi_{f}$, and they show it can be solved in polynomial time. We can use this result to decide whether a final state is universally reachable by a rendez-vous protocol. The idea is to build another rendez-vous protocol where  a process in the final state can send all the processes remained in the inital state to the final state (thanks to a specific message). We have then that the cut-off problem for rendez-vous protocols has a positive answer in this new protocol iff the final state is universally reachable by the original rendez-vous protocol.

  \begin{theorem}\label{thm:rdv-ptime}
Given a rendez-vous protocol $\RvProt=\tuple{Q,\Sigma,q_{in},q_f,E}$, we can decide in \textsc{Ptime} whether $q_f$ is universally reachable by $\RvProt$ or not.
  \end{theorem}

  \begin{proof}
    Let $\RvProt=\tuple{Q,\Sigma,q_{in},q_f,E}$ be a rendez-vous protocol. W.l.o.g we assume that there is no edge in $E$ with destination $q_{in}$ and no edge with source $q_f$, i.e., for all $(q,a,q') \in E$ we have $q'\neq q_{in}$ and $q \neq q_f$ (the protocol being non-deterministic, we can make a copy of $q_{in}$ and $q_f$ respecting this property). From $\RvProt$, we build another rendez-vous protocol $\RvProt'=\tuple{Q,\Sigma \cup \set{end},q_{in},q_f,E'}$ with the same set of states and such that $E'=E \cup \set{(q_{in},?end,q_f),(q_f,!end,q_f)}$. In $\RvProt'$, a process in the final state $q_f$ can send all the processes in $q_{in}$ in the final state $q_f$. We have then that the cut-off problem for rendez-vous networks studied in \cite{balasubramanian-finding-fossacs21} has a positive answer in $\RvProt'$ iff $q_f$ is universally reachable by $\RvProt$.
  \end{proof}

\subsection{No test and many counters}

We show here that when the considered counter protocol has no test, we can translate it into a rendez-vous protocol and use the result of Theorem \ref{thm:rdv-ptime} to show that \TargetP~can be solved in polynomial time. The intuition is that in the constructed rendez-vous protocol, we have some processes simulating the behavior of processes of the counter protocol and other processes simulating the behavior of the counter. Intuitively, each time a process will increment a counter $c$, in the rendez-vous protocol, a rendez-vous request $! \mathtt{inc}(c)$ will be performed and the process answering it to will be sent to a state $1_c$ (as a matter of fact, the number of processes in state $1_c$ will represent the associated counter value). A process in such a state can then answer to a request, doing $?\mathtt{dec}(c)$, to decrement the counter, and in that case it is sent to the final state. A last trick is needed as in \TargetP, we do not require the counter values to be $0$ in the final configuration, so a process in the final state, will be able to request rendez-vous with $!\mathtt{dec}(c)$ to send the remaining processes in $1_c$ to the final state. Not that in order to avoid, a process simulating a counter to play the role of process simulating a process of the counter protocol when reaching the final state, we rely on a new final state $q'_f$ which can be reached by any process reaching the original final state $q_f$ (but cannot be reached by a process simulating a counter).

We move now to the formal definition. Let  $\Prot=\tuple{Q,{k},q_{in},q_f,E}$ be a zero-test free counter protocol. We build the rendez-vous protocol $\RvProt=\tuple{Q',\Sigma,q_{in},q'_f,E'}$ with $Q'=Q \cup \set{1_c \mid c \in [1,k]} \cup \set{q'_f}$ (with $Q \cap \big(\set{1_c \mid c \in [1,k]} \cup \set{q'_f}\big)=\emptyset$) and $\Sigma=\set{\mathtt{inc}(c),\mathtt{dec}(c) \mid c \in [1,k]}$ and $E'$ verifies the following properties:
\begin{itemize}
\item for all $q,q' \in Q$, we have $(q,!\mathtt{inc}(c),q') \in E'$ iff $(q,\mathtt{inc}(c),q') \in E$, and,
\item for all $q,q' \in Q$, we have $(q,!\mathtt{dec}(c),q') \in E'$ iff $(q,\mathtt{dec}(c),q') \in E$, and,
\item for all $q,q' \in Q$, we have  $(q,\tau,q') \in E'$ iff $(q,\tau,q') \in E$, and,
\item $(q_{in},?\mathtt{inc}(c),1_c),(1_c,?\mathtt{dec}(c),q'_f) \in E'$ for all $c \in [1,k]$, and,
\item $(q'_f,!\mathtt{dec}(c),q'_f) \in E'$ for all $c \in [1,k]$,and,
\item $(q_f,\tau,q'_f) \in E$.
\end{itemize}

Relying on the two semantics of counter protocols and rendez-vous protocols, we can easily show that $q_f$ is universally reachable in $\Prot$ if and only if  $q'_f$ universally reachable in $\RvProt$. Thanks to Theorem \ref{thm:rdv-ptime}, we deduce the following result.

 \begin{theorem}\label{thm:target-notest}
\TargetP~ restricted to  zero-test free protocols is in \textsc{Ptime}.
 \end{theorem}

\subsection{Authorizing tests over a single counter}

The translation towards rendez-vous protocols presented in the previous subsection does not work when allowing zero-tests because the communication with rendez-vous does not allow to test the absence of a process ready to communicate through a rendez-vous. However when dealing with a single counter, we can provide another translation relying on the fact that initial and final runs can be reordered into a normal form which can be simulated by rendez-vous protocols. Intuitively such a run will be in a normal form if it can be decomposed into a sequence of subruns, which are of three possible forms: either they consist of a run where a single process is performing some actions and where the  counter value starts at value $0$ and ends at value $0$, or they are of length $2$ and they correspond to an increment performed by a process followed by a decrement performed by another process or they correspond to  a run involving a single process reaching the final control state. This observation allows us to build a rendez-vous protocol where the  runs of the counter protocols performed by a single process between two states will be replaced by an internal actions (we can compute the pairs of states satisfying this property in \textsc{NL} relying on Theorem \ref{thm:fixed-result}) or a rendez-vous between two processes will correspond to a sequence of an increment performed by a process and a decrement performed by another one. We now formalize these intuitions.

In the sequel, we consider a one-counter protocol $\Prot=\tuple{Q,1,q_{in},q_f,E}$. W.l.o.g. we assume that there is no edge in $E$ starting at $q_f$ (i.e., for all $(q,a,q') \in E$, we have $q \neq q_f$). We introduce now some new definitions for the runs of $\Prot$. Let $\rho=\gamma_0 \ltrans{p_1,e_1} \gamma_1 \ltrans{p_2,e_2} \gamma_2 \cdots  \ltrans{p_\ell,e_\ell}\gamma_\ell$ be a $n$-process run of $\Prot$ with $\gamma_i=(\vect{s}_i,v_i)$ for all $i \in [0,\ell]$. We say that:
\begin{itemize}
\item $\rho$ is a \emph{single-active-process run} iff there exists $p \in [1,n]$ such that $p_i=p$ for all $i \in [1,\ell]$ (in that case $p$ is called the \emph{active process}); furthermore it is said to be \emph{final} if $\vect{s}_\ell[p]=q_f$.
\item $\rho$ is a $0$-run [resp. starts with $0$] iff $v_0=v_\ell=0$ [resp. $v_0=0$].
\item $\rho$ stays above $0$ iff $v_i>0$ for all $i \in [1,\ell-1]$.
\item $\rho$ is an inc/dec-run if $\ell=2$ and $p_1\neq p_2$ and $\mathit{lab}(e_1)=\mathtt{inc}$ and $\mathit{lab}(e_2)=\mathtt{dec}$. 
\end{itemize}
Note that if $\rho$ stays above $0$, it means that $\mathit{lab}(e_i) \neq \mathtt{zero}$ for all $i \in [1,\ell]$. Furthermore a run $\rho'=\gamma_0 \ltrans{p'_1,e'_1} \gamma'_1 \ltrans{p'_2,e'_2}
\gamma'_2 \cdots  \ltrans{p'_\ell,e'_\ell}
\gamma_\ell$ is a permutation of a run $\rho=\gamma_0 \ltrans{p_1,e_1} \gamma_1 \ltrans{p_2,e_2}
\gamma_2 \cdots  \ltrans{p_\ell,e_\ell}
\gamma_\ell$ iff there exists a permutation $\pi:[1,\ell] \mapsto [1,\ell]$ such that $(p_i,e_i)=(p'_{\pi(i)},e'_{\pi(i)})$ for all $i \in [1,\ell]$ (the two runs start and end with the same configurations). We say that a $0$-run $\rho$ is in \emph{normal form} iff $\rho=\rho_1.\rho_2\ldots\rho_m$ where for all $i \in [1,m]$, we have that $\rho_i$ is either a single-active-process $0$-run or an inc/dec run.

\begin{lemma}\label{lem:normal-0-run}
If $\rho$ is $0$-run, then there exists a permutation of $\rho$ in normal form.
\end{lemma}

\begin{proof}
  We reason by induction on the length of the run. If $\rho$ is a $0$-run of length $1$, then $\rho$ is in normal formal because it is a single-active-process $0$-run. Assume now that the property holds for all the $0$-runs of length $\ell \geq 1$ and let $\rho$ be a $0$-run of length $\ell+1$. We have two different cases. If $\rho$ does not stay above $0$, then there exists two $0$-runs $\rho_1$ and $\rho_2$ such that $\rho=\rho_1.\rho_2$. By induction hypothesis, there exists $\rho'_1$ a permutation of $\rho_1$ in normal form and $\rho'_2$ a permutation of $\rho_2$ in normal form and the run $\rho'_1.\rho'_2$ corresponds to a permutation of $\rho$ in normal form.

  Assume now that $\rho$ stays above $0$ and that it is not a single-active-process run (otherwise the property holds directly). Then $\rho$ has the following form $\rho=\rho_{inc}.\theta.\rho_{dec}$ where $\rho_{inc}=\gamma \ltrans{p,(q,\mathit{inc},q')} \gamma'$, $\rho_{dec}=\gamma'' \ltrans{p',(q'',\mathit{dec},q''')} \gamma'''$ and $\theta=\gamma_0 \ltrans{p_1,e_1} \gamma_1 \ltrans{p_2,e_2} \gamma_2 \cdots  \ltrans{p_\ell,e_\ell} \gamma_{\ell}$ with $\gamma_0=\gamma'$, $\gamma_\ell=\gamma''$ and for all $i \in [1,\ell]$, we have $\mathit{lab}(e_i)\neq \mathtt{zero}$ and $\gamma_{i}=(\vect{s}_i,v_i)$ such that $v_0=v_\ell=1$ and $v_i\geq 1$.

Before we continue, we deal with the specific case where $\theta$ is a single-active-process run with $p$ as an active process. As $\rho$ is not an single-active-process run, we have necessarily $p'\neq p$. In that case, we can move the actions of $\rho_{dec}$ just after $\rho_{inc}$ to obtain a $0$-run of the form $\rho_{inc}.\rho'_{dec}.\theta'$ with $\mathit{actions}(\rho'_{dec})=\mathit{actions}(\rho_{dec})=(p',(q'',\mathit{dec},q'''))$ and $\mathit{actions}(\theta')=\theta)$. In that case, note that applying the induction hypothesis to $\theta'$, we obtain a run $\theta''$ that is a permutation of $\theta'$ in normal form and the run  $\rho_{inc}.\rho'_{dec}.\theta''$ is a permutation of $\rho$ in normal form.

We suppose now that $\theta$ is not a single-active-process run with $p$ as an active process. From $\theta$, we can build a $0$-run $\theta'$ which follows the same step as $\theta$ starting at a counter value equal to $0$ (instead of $1$). Formally, $\theta'=\gamma'_0 \ltrans{p_1,e_1} \gamma'_1 \ltrans{p_2,e_2} \gamma'_2 \cdots  \ltrans{p_\ell,e_\ell} \gamma'_{\ell}$ where for all $i \in [1,n]$ we have $\gamma'_i=(\vect{s}_i,v'_i)$ with $v'_i=v_i-1$. Note that by construction $\mathit{actions}(\theta')=\mathit{actions}(\theta)$. We can now apply the induction hypothesis to obtain a run $\theta''$ that is a permutation of the run $\theta'$ in normal form.  Hence $\theta''=\theta_1.\theta_2\ldots \theta_m$ where  $\theta_i$ is either a single-active process $0$-run or an inc/dec run for all $i \in [1,m]$.  Now take $k$ to be the smallest index in $[1,m]$ such that $\theta_k$ is not a single-active-process $0$-run with $p$ as an active process (it necessarily exists as $\theta$ is not a single-active-process run with $p$ as an active process). We proceed with a case analysis:
  \begin{itemize}
  \item $\theta_k$ is a single-active-process $0$-run with $p'$ as an active process (and $p' \neq p$). In that case, from $\rho=\rho_{inc}.\theta.\rho_{dec}$ we can build another $0$-run of the form $\theta'_k.\rho'_{inc}.\theta'_1.\theta'_2\ldots\theta'_{k-1}.\theta'_{k+1}\linebreak[0]\ldots\linebreak[0]\theta'_m.\rho_{dec}$ where $\mathit{actions}(\rho'_{inc})=\mathit{action}(\rho_{inc})$ and $\mathit{actions}(\theta'_i)=\mathit{actions}(\theta_i)$ for all $i \in [1,m]$. This is possible as in $\theta''$ before the index $k$, the active process $p'$ of $\theta_k$ which is different than $p$ did not perform any actions, this is why we can move its actions which have globally no effect on the counter value at the beginning of run $\rho$. We have that $\theta'_k$ is a single-active-process $0$-run and by applying the induction hypothesis on the $0$-run $\rho'_{inc}.\theta'_1.\theta'_2\ldots\theta'_{k-1}.\theta'_{k+1}\ldots\theta'_m.\rho_{dec}$, we obtain a run permutation of $\rho$ in normal form.
 \item $\theta_k$ is an inc/dec run with $\mathit{actions}(\theta_k)=(p_{inc},e_{inc})(p_{dec},e_{dec})$. In that case, we can exchange the order of the two actions to obtain a run $\rho'=\rho_{inc}.\theta'_1.\theta'_2\ldots\theta'_{k-1}.\linebreak[0]\theta^{(1)}_k.\theta^{(2)}_k\theta'_{k+1}\ldots\linebreak[0]\theta'_m.\rho_{dec}$ where $\mathit{actions}(\rho'_{inc})=\mathit{action}(\rho_{inc})$ and $\mathit{actions}(\theta'_i)=\mathit{actions}(\theta_i)$ for all $i \in [1,m] \setminus \set{k}$ and $\mathit{actions}(\theta^{(1)}_k)=(p_{dec},e_{dec})$ and $\mathit{actions}(\theta^{(2)}_k)(p_{inc},e_{inc})$. This exchange is possible because in the run $\theta$ the counter never goes below $1$.  The runs $\rho_{inc}.\theta'_1.\theta'_2\ldots\theta'_{k-1}.\theta^{(1)}_k$ and $\theta^{(2)}_k\theta'_{k+1}\ldots\theta'_m.\rho_{dec}$ are then $0$-runs, which can both be put in normal form by induction hypothesis to obtain a run that is a permutation of $\rho$ in normal form.
  \end{itemize}
\end{proof}

Relying on this we can extend our definition of runs in normal forms to initial runs bringing every process in the final state $q_f$. We say that an initial $n$-process run $\rho$ such that $\mathit{last}(\rho) \in \Gamma_{\forall f}$ is in normal formal  iff $\rho=\rho_1.\rho_2\ldots\rho_m$ where for all $i \in [1,m]$, we have that $\rho_i$ is either a single-active-process $0$-run or an inc/dec run or a final single-active-process run with active process $p$ such that $\mathit{path}^{(p)}(\rho_i)$ is a path feasible with $1$ process. Relying one the previous lemma and on a similar reasoning, we can prove the following result.

\begin{lemma}\label{lem:normal-form}
  If $\rho$ is a run starting with $0$ such that $\mathit{last}(\rho) \in \Gamma_{\forall f}$, then there exists a permutation of $\rho$ in normal form.
\end{lemma}

\begin{proof}
 Consider a run $\rho$ starting with $0$ and such that $\mathit{last}(\rho) \in \Gamma_{\forall f}$. Here again, as for the proof of Lemma \ref{lem:normal-0-run}, we can reason by induction on the length of $\rho$. If it is of length $1$, then it is a single-active-process run and the path of the active process is a final path feasible with $1$ process. If $\rho$ does not stay above $0$, we can cut it into two runs: the first one is a $0$-run for which we can apply Lemma \ref{lem:normal-0-run} and the second one on which we apply the induction hypothesis. If $\rho$ stays above $0$, then its first action is an increment and as for the proof of Lemma \ref{lem:normal-0-run}, we can consider the $0$-run $\rho'$ build from $\rho$ by deleting this first action and following them the same step. By induction hypothesis there exists a permutation $\rho''$ of $\rho'$ which is in normal formal. If in $\rho''$ there is no inc/dec run, then all the processes involved in $\rho''$ can reach $q_f$ on their own and replacing the process performing the increment in $\rho$ at the beginning of $\rho''$ we obtain again a run where all processes can reach on their own $q_f$. If in $\rho''$ there is an inc/dec run, then we can exchange the order locally between the increment and decrement and we obtain again a $0$-run followed by a run bringing everyone to $q_f$, both of them can be put in normal form applying Lemma \ref{lem:normal-0-run} and the induction hypothesis.
\end{proof}

This lemma allows us to build from the counter protocol $\Prot$ a rendez-vous protocol $\RvProt$ which simulates the behavior of a run in normal form. We have indeed the following cases to deal with. Either each process can perform actions on their own going from a state $q$ to $q'$ leaving the counter equal to $0$, in that case we can compute such pairs of states in $\Prot$ and simulate their path by an internal action on $\RvProt$. Or one process performs an increment followed by a decrement of another process; this can be simulated by a rendez-vous (note that afterwards, if the counter was equal to $0$ before, it stays equal after). Or a process can go on its own to $q_f$ (which is by hypothesis a deadlock state) and this is again simulated by an internal action in  $\RvProt$. Formally, we rely on the two following sets of states: $S_{0\rightarrow^\ast0}=\set{(q,q') \in Q^2 \mid ([q],[0]) \rightarrow^\ast ([q'],[0])}$  (pair of states from which there exists a $1$-process $0$-run starting in the first state and ending in the second state) and $S_{0\rightarrow^\ast f}=\set{q \in Q \mid ([q],[0]) \rightarrow^\ast ([q_f],[v]) \mbox{for some } v \in \nat}$ (states from which there is a $1$-process run reaching the final state). Thanks to Theorem \ref{thm:fixed-result}, these two sets can be computed in polynomial time. We build the rendez-vous protocol $\RvProt=(Q,\Sigma,q_{in},q_f,E')$ which has the same set of states as $\Prot$, where the alphabet of messages $\Sigma$ is such that $\Sigma={\mathtt{i-d}}$ and where $E'$ verifies:
\begin{itemize}
\item $(q,\tau,q') \in E'$ iff $(q,q') \in S_{0\rightarrow^\ast0}$ or ($q'=q_f$ and $q \in S_{0\rightarrow^\ast f}$), and,
\item $(q,!\mathtt{i-d},q') \in E'$ iff $(q,\mathtt{inc},q') \in E$, and,
\item $(q,?\mathtt{i-d},q') \in E'$ iff $(q,\mathtt{dec},q') \in E$.
\end{itemize}

Now thanks to Lemma \ref{lem:normal-form} and relying on the semantics of the two protocols, we have that $q_f$ is universally reachable by $\Prot$ iff $q_f$ is universally reachable by $\RvProt$. Relying on Theorem \ref{thm:rdv-ptime} and on the fact that $\RvProt$ can be computed in polynomial time (as we have explained before), this leads us to the last result of this section.

\begin{theorem}\label{thm:target-one}
\TargetP~restricted to  one-counter protocols is in \textsc{Ptime}.
 \end{theorem}

\bibliography{biblio.bib}

@book{minsky-computation-67,
  author = 	 {M. Minsky},
  title = 	 {Computation, Finite and Infinite Machines},
  publisher = 	 {Prentice Hall},
  year = 	 {1967},
}

@book{lipton-reach-76,
  title={The reachability problem requires exponential space},
  author={Lipton, R.J.},
  series={Research report (Yale University. Department of Computer Science)},
  year={1976},
  publisher={Department of Computer Science, Yale University}
}

@article{rackoff-covering-78,
  author    = {C. Rackoff},
  title     = {The Covering and Boundedness Problems for Vector Addition Systems},
  journal   = {Theoretical Computer Science},
  volume    = {6},
  pages     = {223-231},
  year      = {1978}
}

@inproceedings{lafourcade-intruder-rta05,
  author    = {P. Lafourcade and
               D. Lugiez and
               R. Treinen},
  title     = {Intruder Deduction for \emph{AC}-Like Equational Theories with Homomorphisms},
  booktitle = {{RTA}'05},
  series    = {Lecture Notes in Computer Science},
  volume    = {3467},
  pages     = {308-322},
  publisher = {Springer},
  year      = {2005}
}

@article{chistikov-shortest-lmcs19,
  author    = {D. Chistikov and
               W. Czerwinski and
               P. Hofman and
               M. Pilipczuk and
               M. Wehar},
  title     = {Shortest paths in one-counter systems},
  journal   = {Logical Methods in Computer Science},
  volume    = {15},
  number    = {1},
  year      = {2019}
}

@article{ladner-circuit-sigact75,
  author    = {Richard E. Ladner},
  title     = {The circuit value problem is log space complete for \emph{P}},
  journal   = {{SIGACT} News},
  volume    = {7},
  number    = {1},
  pages     = {18--20},
  year      = {1975}
}

@article{german-reasoning92,
	author    = {S. M. German and
	A. Prasad Sistla},
	title     = {Reasoning about Systems with Many Processes},
	journal   = {Journal of the {ACM}},
	volume    = {39},
	number    = {3},
	pages     = {675--735},
	year      = {1992}
}

@inproceedings{balasubramanian-finding-fossacs21,
	author    = {A. R. Balasubramanian and
	J. Esparza and
	M. Raskin},
	title     = {Finding Cut-Offs in Leaderless Rendez-Vous Protocols is Easy},
	booktitle = {{FOSSACS}'21},
	series    = {LNCS},
	volume    = {12650},
	pages     = {42--61},
	publisher = {Springer},
	year      = {2021},
}

@inproceedings{horn-deciding-concur20,
	author    = {F. Horn and
	A. Sangnier},
	title     = {Deciding the Existence of Cut-Off in Parameterized Rendez-Vous Networks},
	booktitle = {{CONCUR}'20},
	series    = {LIPIcs},
	volume    = {171},
	pages     = {46:1--46:16},
	publisher = {Schloss Dagstuhl - Leibniz-Zentrum f{\"{u}}r Informatik},
	year      = {2020},
}

@article{aminof-parameterized-dc18,
  author       = {B. Aminof and
                  T. Kotek and
                  S. Rubin and
                  F. Spegni and
                  H. Veith},
  title        = {Parameterized model checking of rendezvous systems},
  journal      = {Distributed Comput.},
  volume       = {31},
  number       = {3},
  pages        = {187--222},
  year         = {2018},
  bibsource    = {dblp computer science bibliography, https://dblp.org}
}

@article{durand-model-fmsd17,
  author       = {A. Durand{-}Gasselin and
                  J. Esparza and
                  P. Ganty and
                  R. Majumdar},
  title        = {Model checking parameterized asynchronous shared-memory systems},
  journal      = {Formal Methods Syst. Des.},
  volume       = {50},
  number       = {2-3},
  pages        = {140--167},
  year         = {2017}
}

@article{esparza-parameterized-acm16,
  author       = {J. Esparza and
                  P. Ganty and
                  R. Majumdar},
  title        = {Parameterized Verification of Asynchronous Shared-Memory Systems},
  journal      = {J. {ACM}},
  volume       = {63},
  number       = {1},
  pages        = {10:1--10:48},
  year         = {2016}
}

@inproceedings{aminof-parameterized-vmcai14,
  author       = {B. Aminof and
                  S. Jacobs and
                  A. Khalimov and
                  S. Rubin},
  title        = {Parameterized Model Checking of Token-Passing Systems},
  booktitle    = {{VMCAI}'14},
  series       = {Lecture Notes in Computer Science},
  pages        = {262--281},
  publisher    = {Springer},
  year         = {2014}
}

@inproceedings{esparza-verification-lics99,
  author       = {J. Esparza and
                  A. Finkel and
                  R. Mayr},
  title        = {On the Verification of Broadcast Protocols},
  booktitle    = {{LICS}'99},
  pages        = {352--359},
  publisher    = {{IEEE} Computer Society},
  year         = {1999}
}

@inproceedings{delzanno-complexity-fsttcs12,
  author       = {G. Delzanno and
                  A. Sangnier and
                  R. Traverso and
                  G. Zavattaro},
  title        = {On the Complexity of Parameterized Reachability in Reconfigurable
                  Broadcast Networks},
  booktitle    = {{FSTTCS}'12},
  series       = {LIPIcs},
  pages        = {289--300},
  publisher    = {Schloss Dagstuhl - Leibniz-Zentrum f{\"{u}}r Informatik},
  year         = {2012}
}

@inproceedings{delzanno-complexity-concur10,
  author       = {G. Delzanno and
                  A. Sangnier and
                  G. Zavattaro},
  title        = {Parameterized Verification of Ad Hoc Networks},
  booktitle    = {{CONCUR}'10 },
  series       = {Lecture Notes in Computer Science},
  pages        = {313--327},
  publisher    = {Springer},
  year         = {2010}
}

@inproceedings{mascle-model-concur23,
  author       = {C. Mascle and
                  A. Muscholl and
                  I. Walukiewicz},
  title        = {Model-Checking Parametric Lock-Sharing Systems Against Regular Constraints},
  booktitle    = {{CONCUR}'23},
  series       = {LIPIcs},
  pages        = {24:1--24:17},
  publisher    = {Schloss Dagstuhl - Leibniz-Zentrum f{\"{u}}r Informatik},
  year         = {2023}
}

@inproceedings{guillou-safety-concur23,
  author       = {L. Guillou and
                  A. Sangnier and
                  N. Sznajder},
  title        = {Safety Analysis of Parameterised Networks with Non-Blocking Rendez-Vous},
  booktitle    = {{CONCUR}'23},
  series       = {LIPIcs},
  pages        = {7:1--7:17},
  publisher    = {Schloss Dagstuhl - Leibniz-Zentrum f{\"{u}}r Informatik},
  year         = {2023}
 }

@inproceedings{delzanno-towards-tacas02,
  author       = {G. Delzanno and
                  J.-F. Raskin and
                  L. Van Begin},
  title        = {Towards the Automated Verification of Multithreaded Java Programs},
  booktitle    = {{TACAS}'02},
  series       = {Lecture Notes in Computer Science},
  pages        = {173--187},
  publisher    = {Springer},
  year         = {2002}
}

@inproceedings{haase-reachability-concur09,
  author       = {C. Haase and
                  S. Kreutzer and
                  J. Ouaknine and
                  J. Worrell},
  title        = {Reachability in Succinct and Parametric One-Counter Automata},
  booktitle    = {{CONCUR}'09},
  series       = {Lecture Notes in Computer Science},
  pages        = {369--383},
  publisher    = {Springer},
  year         = {2009}
}

@inproceedings{goller-model-icalp10,
  author       = {S. G{\"{o}}ller and
                  C. Haase and
                  J. Ouaknine and
                  J. Worrell},
  title        = {Model Checking Succinct and Parametric One-Counter Automata},
  booktitle    = {{ICALP}'10},
  series       = {Lecture Notes in Computer Science},
  pages        = {575--586},
  publisher    = {Springer},
  year         = {2010}
}

@inproceedings{leroux-reachability-lics19,
	author       = {J. Leroux and
	S. Schmitz},
	title        = {Reachability in Vector Addition Systems is Primitive-Recursive in
	Fixed Dimension},
	booktitle    = {{LICS}'19},
	pages        = {1--13},
	publisher    = {{IEEE}},
	year         = {2019}
}

@inproceedings{leroux-reachability-focs21,
	author       = {J. Leroux},
	title        = {The Reachability Problem for Petri Nets is Not Primitive Recursive},
	booktitle    = {{FOCS}'21},
	pages        = {1241--1252},
	publisher    = {{IEEE}},
	year         = {2021}
}

@inproceedings{czerwinski-reachability-focs21,
	author       = {W. Czerwinski and
	L. Orlikowski},
	title        = {Reachability in Vector Addition Systems is Ackermann-complete},
	booktitle    = {{FOCS}'21},
	pages        = {1229--1240},
	publisher    = {{IEEE}},
	year         = {2021}
}

@inproceedings{chistikov-tractability-focs24,
  author       = {D. Chistikov and
                  W. Czerwinski and
                  F. Mazowiecki and
                  L- Orlikowski and
                  H. Sinclair{-}Banks and
                  K. W{\k{e}}grzycki},
  title        = {The Tractability Border of Reachability in Simple Vector Addition
                  Systems with States},
  booktitle    = {{FOCS}'24},
  pages        = {1332--1354},
  publisher    = {{IEEE}},
  year         = {2024}
}

@article{blondin-reachability-acm21,
  author       = {M. Blondin and
                  M. Englert and
                  A. Finkel and
                  S. G{\"{o}}ller and
                  C. Haase and
                  R. Lazic and
                  P. McKenzie and
                  P. Totzke},
  title        = {The Reachability Problem for Two-Dimensional Vector Addition Systems
                  with States},
  journal      = {J. {ACM}},
  volume       = {68},
  number       = {5},
  pages        = {34:1--34:43},
  year         = {2021}
}

@inproceedings{englert-reachability-lics16,
  author       = {M. Englert and
                  R. Lazic and
                  P. Totzke},
  title        = {Reachability in Two-Dimensional Unary Vector Addition Systems with
                  States is NL-Complete},
  booktitle    = {{LICS} '16},
  pages        = {477--484},
  publisher    = {{ACM}},
  year         = {2016}
}

@inproceedings{kukovec-reachability-concur18,
  author       = {J. Kukovec and
                  I. Konnov and
                  J. Widder},
  title        = {Reachability in Parameterized Systems: All Flavors of Threshold Automata},
  booktitle    = {{CONCUR}'18},
  series       = {LIPIcs},
  volume       = {118},
  pages        = {19:1--19:17},
  publisher    = {Schloss Dagstuhl - Leibniz-Zentrum f{\"{u}}r Informatik},
  year         = {2018}
}

@inproceedings{balasubramanian-complexity-atva20,
  author       = {A. R. Balasubramanian and
                  J. Esparza and
                  M. Lazic},
  title        = {Complexity of Verification and Synthesis of Threshold Automata},
  booktitle    = {{ATVA}'20},
  series       = {LNCS},
  volume       = {12302},
  pages        = {144--160},
  publisher    = {Springer},
  year         = {2020}
}


\end{document}